\documentstyle[aps,prb,epsf]{revtex} 
\begin{document} 
%\draft 

\title{Interaction effects in non-Hermitian  
models of vortex physics} 
\author{Kihong Kim\cite{kk}}
\address{Department of Molecular Science
and Technology, Ajou University, Suwon 442-749, Korea} 
\author{David R. Nelson\cite{drn}} 
\address{Lyman Laboratory of Physics, Harvard University, 
Cambridge, Massachusetts 02138} 

\wideabs{
\maketitle  
\begin{abstract} 
Vortex lines in superconductors in an external magnetic field
slightly tilted from randomly-distributed parallel columnar defects
can be modeled by a system of interacting bosons in a 
non-Hermitian vector potential and a random scalar potential.
We develop a theory of 
the strongly-disordered non-Hermitian boson Hubbard model
using the Hartree-Bogoliubov approximation and apply it to calculate
the complex energy spectra, the vortex tilt angle and the tilt modulus of 
($1+1$)-dimensional directed flux line systems. 
We construct the phase diagram associated with the 
flux-liquid to Bose-glass transition and 
find that, close to the phase boundary, the tilted
flux liquid phase is characterized by a band of localized
excitations, with two mobility edges in its low-energy spectrum.  
\end{abstract} 
}

\section{Introduction} 
 
An external magnetic field between the lower
critical field $H_{c1}$ 
and the upper critical field $H_{c2}$ 
of a type-II superconductor 
penetrates the material in the form of flexible flux 
lines (or vortices).  
In high-temperature cuprate superconductors, which are 
strongly type-II superconductors,  
a wide variety of thermodynamic and dynamic 
phases can be formed in a system of magnetic flux 
lines by an interplay of strong thermal 
fluctuations, intervortex interactions, material anisotropy,  
and disordering and pinning effects 
generated by various kinds 
of defects and impurities.\cite{blat} 
In addition to the famous Abrikosov vortex lattice phase, 
a number of fundamentally new liquid, crystalline and  
glassy phases have been proposed over the last  
decade.\cite{nels,mfish,vino,bragg} 
The study of these phases and the transitions  
among them is a subject of active theoretical and 
experimental research in condensed matter physics. 
 
A convenient way of 
understanding the system of interacting flux lines is 
provided by the formal mapping between the classical 
statistical mechanics of 
($d+1$)-dimensional directed flux lines 
and the nonrelativistic quantum mechanics 
of $d$-dimensional bosons.\cite{nels,dhlee} 
In this mapping, the flux lines traversing the sample 
along the direction of the external magnetic field,  
${\bf H}=H\hat{\bf z}$, correspond to the boson 
world lines propagating through the {\it imaginary}
time dimension. The thickness of the sample 
in the $z$ direction, $L_z$,  
is analogous to the inverse temperature 
$\beta\hbar$ ($=\hbar/k_BT$), which is the size 
of the imaginary time dimension, $L_\tau$.  
The effective line 
tension $\tilde\epsilon_1$ corresponds to the boson mass $m$ 
and thermal fluctuations proportional to $k_BT$ 
play the role of quantum fluctuations proportional to 
$\hbar$. When the flux lines satisfy periodic 
boundary conditions in the $z$ direction or in the  
thermodynamic limit $L_z\rightarrow\infty$ 
(corresponding to the zero temperature limit
for the fictitious bosons), 
the boson-vortex mapping becomes precise. 
When free boundary conditions are more appropriate to the flux line 
system, the boson 
mapping must be corrected by finite size effects.\cite{taub} 
More details of the boson-vortex mapping will be discussed in  
Sec.~\ref{sec:model} and summarized in Table~I. 
 
Using the mapping between flux lines and bosons, 
it has been suggested that, in clean high-temperature 
superconductors, the Abrikosov vortex lattice can melt 
and be transformed to an entangled flux liquid, which is 
an analogue of the bosonic superfluid phase.\cite{nels} 
It has also been argued that  
it is possible to produce new glassy phases by adding various 
kinds of defects to clean superconductors.\cite{mfish,vino,bragg} 
In the present work, 
we focus on the ``Bose glass'' phase, which can be produced,
for instance, by  
artificially creating parallel 
columnar defects inside a superconductor using heavy ion 
irradiation.\cite{vino,bud,civ} 
These defects are assumed to be distributed randomly 
in the plane perpendicular to the columns. 
In the spirit of the boson-vortex mapping, 
the columnar defects provide a static random potential for 
bosons, if they lie parallel to the external magnetic 
field. Interacting bosons in a strong static 
random potential can be localized and  
form the insulating Bose glass phase. The quantum superfluid to Bose glass 
transition, occurring when the effective disorder strength is changed  
across a critical value at zero temperature, has been 
studied extensively 
during the last decade.\cite{hertz,mhl,giam,dfish,fwgf,kim} 
The knowledge obtained from 
this research can be directly applied to the flux-liquid to Bose-glass 
transition in high-temperature superconductors 
with parallel columnar defects.       
   
If the direction of the external 
magnetic field does {\it not} coincide with 
that of columnar defects, it is convenient to separate the transverse 
component of the field ${\bf H}_\perp$ from the parallel component 
${\bf H}_\parallel$ ($=H_\parallel\hat{\bf z}$).  
When $H_\perp\ll H_\parallel$, or equivalently when the tilt angle 
of the external field, 
$\theta_H=\tan^{-1}(H_\perp/H_\parallel)$, 
is sufficiently small, one can show that the $z$ axis  
corresponds to the imaginary time dimension and the transverse 
component $H_\perp$ plays the role of an {\it imaginary} vector potential 
for bosons.\cite{dous} 
The corresponding fictitious quantum Hamiltonian 
is {\it non-Hermitian} and in general possesses 
{\it complex} energy eigenvalues. 
 
Recently, there has been great interest 
in non-Hermitian quantum mechanics 
of {\it noninteracting} bosons
in a constant imaginary vector potential, which we call ${\bf g}$, 
and a random scalar potential.\cite{hat,efet,brou,mudr}
In the boson-vortex analogy, 
${\bf g}$ is proportional to ${\bf H}_\perp$. 
Since the Hamiltonian is  
non-Hermitian, the energy eigenvalues can be either real or complex. 
Hatano and Nelson have argued that all states with complex eigenvalues 
are extended, whereas those with real eigenvalues are either localized 
or extended.\cite{hat} 
It is simpler to describe the basic phenomena in the  
($1+1$)-dimensional case. A model system (with periodic boundary
conditions) appropriate to flux lines in ($1+1$) dimensions is 
sketched in Fig.~\ref{fig:pbc}.\cite{hat} 
In the absence of the  
imaginary vector potential, it is well-known that  
all eigenstates are localized (in both one and two dimensions) 
and the corresponding eigenvalues are real. 
This continues to be the case if ${\bf g}$   
is sufficiently weak   
or the random potential is sufficiently strong. 
As the imaginary vector potential, or the transverse magnetic field, 
is increased above a critical value $g_1$, there appear a few extended  
states with complex eigenvalues in the center of the energy band. 
The spectrum of 
this group of extended states is shaped 
like a bubble on a complex energy plane  
and the two points where the bubble is attached 
to two real lines of localized states  
are the mobility edges separating 
the extended states from the localized ones (see Fig.~\ref{fig:hsp}). 
As $g$ increases further, more states change from 
being localized to extended and the complex bubble expands. Ultimately, 
above another critical value $g_2$ of the imaginary vector potential,  
all states including the ground state
become extended and the entire spectrum is shaped like 
a single complex bubble. The basic physics is similar in two 
dimensions, although localized and extended states coexist near 
the band center and 
the spectrum is much more complicated.\cite{hat}
 
These results suggest the existence of  
a novel delocalization transition even in 
one- and two-dimensional non-Hermitian systems. 
In a small external transverse magnetic field, 
the flux lines remain in localized states and are strongly
pinned by columnar defects. 
Since the system displays perfect diamagnetism
against the transverse field in this case, 
this phenomenon has been termed the {\it transverse Meissner
effect}.
However, above a certain critical transverse 
field, the flux lines are depinned from columnar defects 
and form a tilted flux liquid. A schematic phase diagram  
of high-temperature superconductors in the $T$-$H_\perp$ plane 
is shown in Fig.~\ref{fig:spd}. 
 
In spite of the large number of papers 
devoted to the non-Hermitian problem, 
there has been very little research on the effects of interactions 
on the delocalization phenomena in non-Hermitian systems. 
One may attempt to take account of the interaction effects  
by assuming that the nature of the spectrum in the interacting 
case remains the same as in the noninteracting one 
and forbidding multiple occupancy of the single-particle
localized eigenstates. 
Let us consider the ($1+1$)-dimensional case and 
assume that the transverse field $H_\perp$ is 
in a range such that there is a mobility edge separating  
low-energy localized states from high-energy extended ones.
In the boson-vortex mapping, 
the parallel component of the field, $H_\parallel$, 
corresponds to the chemical potential $\mu$ 
of the bosonic system. 
As $H_\parallel$ is increased at a fixed $H_\perp$, the states
are filled  
in order of increasing energy to obtain the ground state. 
At a critical value of $H_\parallel$, 
all states below the mobility edge are filled 
and additional flux lines begin to fill the extended states 
just above the boundary. These extended states describe   
tilted flux lines and the tilt angle is known to be finite 
at the mobility edge.\cite{hat}
The Hartree-Bogoliubov results presented here suggest that   
this naive picture, which is similar
to the ``inert layer'' picture used to explain
the superfluid onset transition in $^4$He adsorbed in porous media or
on random substrates,\cite{rep,chan,fino} is wrong and needs to 
be substantially modified. 
    
Hwa {\it et al.} have developed 
a physical picture of the delocalized 
phase and the delocalization transition  
in ($1+1$) dimensions in the presence of interactions.\cite{hwa}
When $H_\perp$ is larger than a critical value,
each flux line jumps from one columnar defect
to another via kinks. These kinks line up and form chains
to reduce the interaction energy cost. 
The density of the chains goes continuously to zero 
as $H_\perp$ decreases to the critical value 
and the universality class of the delocalization transition  
is predicted to be the same as that of the standard 
commensurate-incommensurate transition in ($1+1$) dimensions. 
The physical picture in the ($2+1$)-dimensional case 
is less clear and expected to be more complicated 
than that in the ($1+1$)-dimensional one.  
More recently, Lehrer and Nelson studied 
the non-Hermitian Mott-insulator to flux-liquid  
transition occurring in a system of interacting  
flux lines with a {\it periodic} array of columnar defects, 
using both the mean-field theory and the renormalization-group 
method.\cite{lehr} The generic universality class of 
the non-Hermitian transition was found to be  
the same as that of flux lines entering the Meissner phase at 
the lower critical field $H_{c1}$.  

The main aim of this work is  
to develop a quantitative theory 
for understanding the delocalized phase 
and the delocalization transition in 
{\it interacting} non-Hermitian systems
in the presence of {\it strong} disorder. 
We apply the theory to the disordered flux liquid phase
and to the flux-liquid to Bose-glass transition
in high-temperature superconductors with columnar defects. 
Generalizing the model of Ref.~17 
to the short-range-interacting case, 
we introduce a $d$-dimensional non-Hermitian Hubbard 
model for {\it lattice} bosons in a  
random potential: 
\begin{eqnarray} 
{\mathcal H}&=&-\frac{t}{2}\sum_{\bf x}\sum_{\nu=1}^d\left( 
e^{{\bf g}\cdot{\bf e}_\nu/\hbar} 
a_{{\bf x}+{\bf e}_\nu}^\dagger a_{\bf x} 
+e^{-{\bf g}\cdot{\bf e}_\nu/\hbar} 
a_{\bf x}^\dagger a_{{\bf x}+{\bf e}_\nu} 
\right) 
\nonumber\\ 
&&+\sum_{\bf x}V_{\bf x}a_{\bf x}^\dagger 
a_{\bf x}+\frac{U}{2}\sum_{\bf x}a_{\bf x}^\dagger 
a_{\bf x}^\dagger a_{\bf x}a_{\bf x}, 
\label{eq:bhhlb} 
\end{eqnarray}
where $a_{\bf x}^\dagger$ and $a_{\bf x}$ are boson creation and  
annihilation operators at lattice site ${\bf x}$ respectively. 
The hopping integral $t$ sets the scale of the boson ``kinetic energy'' 
and $U$ is the strength of the repulsive 
on-site interaction.   
The random potential $V_{\bf x}$ is assumed to be   
uncorrelated in space and uniformly distributed 
between $-\Delta$ and $\Delta$. 
The constant vector ${\bf g}$ is a non-Hermitian external field and  
the vectors $\{{\bf e}_\nu\}$ are the unit lattice vectors.  
We assume a hypercubic lattice with the 
lattice constant $a$, therefore, $\vert{\bf e}_\nu\vert=a$. 
It is convenient to introduce a {\it dimensionless}
non-Hermitian field ${\bf h}\equiv {\bf g}a/\hbar$.   
The connection between this model and the physics of vortices 
will be discussed in more detail in Sec.~\ref{sec:model}. 
The transverse component of the 
external magnetic field ${\bf H}_\perp$ 
is related to the non-Hermitian field ${\bf h}$ via 
\begin{equation} 
{\bf h}=\frac{\phi_0{\bf H}_\perp a}{4\pi k_BT}, 
\end{equation} 
where $\phi_0=\pi\hbar c/e$ is the magnetic flux quantum.
We use boson notation in the discussion below.

The Hermitian version of the boson Hubbard model (\ref{eq:bhhlb}),
which is obtained by setting ${\bf g}={\bf 0}$, and its continuum 
counterpart
have been studied extensively using a variety of methods,
which include numerical simulations, scaling analysis,
renormalization group calculations
and the Hartree-Bogoliubov 
approximation.
\cite{hertz,mhl,giam,dfish,fwgf,kim,cepe,sca,zhang,gunn,rokh,huang}  
Inspired by an earlier work by Lee and Gunn,\cite{gunn}  
Singh and Rokhsar developed a   
numerical method of studying the disordered superfluid phase  
and the quantum superfluid to Bose-glass 
transition in a system of interacting 
bosons in a random potential,\cite{rokh} 
for the purpose of understanding some experiments done with 
$^4$He adsorbed in porous media.\cite{rep,chan,fino}  
Using a combination of the Hartree approximation 
and the Bogoliubov theory, they calculated 
the condensate depletion and the superfluid density, 
as well as the excitation spectrum and the density of states.
 
In the present work,  
we generalize Singh and Rokhsar's 
method to the non-Hermitian case.  
The first step is to determine the ground state wave function 
in a self-consistent manner using the Hartree approximation. 
We find that the random potential is screened by a (Hartree) 
repulsive interaction with other particles in the ground 
state. The screened potential is weaker and smoother than the  
original random potential. In the next step, we apply the 
Bogoliubov canonical transformation \cite{bogol} to the ground 
and excited states obtained using the Hartree approximation 
and derive the Bogoliubov excitation spectrum and the  
ground state energy. By taking derivatives of the ground  
state energy $E_{g}$ with respect to the non-Hermitian field, 
we calculate the quantities such as 
the imaginary current ${\bf J}_I$ and the superfluid density $n_s$.

The imaginary current in the ground state
is defined by
\begin{equation} 
{\bf J}_I=-\frac{\partial}{\partial {\bf g}}\left(\frac{E_g}{N}\right), 
\label{eq:ic}
\end{equation}
where $N$ is the number of bosons (or flux lines).\cite{hat}
This quantity corresponds to the transverse component of 
the magnetic induction, ${\bf B}_\perp$, 
in flux line systems.
The vortex tilt angle, 
\begin{equation}
\theta_v=\tan^{-1}
\left(\frac{B_\perp}{B_\parallel}\right),
\end{equation}
is related to $J_I$ by
\begin{eqnarray}
\tan\theta_v&=&
-\frac{\partial}{\partial h}\left(\frac{E_g}{tN}\right)
\equiv \frac{J_I}{J_0},~~J_0=\frac{ta}{\hbar}.
\label{eq:tilta}
\end{eqnarray}
The quantity $J_0$ can be expressed as $\hbar a^{-1}/m$,
since, in the continuum limit, the hopping integral
$t$ is related to the boson mass $m$ (this is 
the tilt modulus of the vortices) by 
\begin{equation}
m=\frac{\hbar^2}{ta^2}.
\end{equation}

In a system of volume ${\mathcal V}$, the superfluid density
is defined by 
\begin{eqnarray} 
n_s&=&-\frac{\hbar^2}{ta^2}\lim_{\delta g\rightarrow 0}
\frac{\partial^2}{\partial (\delta g)^2}
\left[\frac{E_g\left({\bf g}+\delta{\bf g}\right)}
{{\mathcal V}}\right]
\nonumber\\
&=&-\lim_{\delta h\rightarrow 0}
\frac{\partial^2}{\partial (\delta h)^2}
\left[\frac{E_g\left({\bf h}+\delta{\bf h}\right)}
{t{\mathcal V}}\right], 
\label{eq:sfd}
\end{eqnarray} 
where $\delta {\bf h}\equiv (a/\hbar)\delta {\bf g}$.
The direction of the vector $\delta{\bf g}$ is that of the superflow.
In general, the superfluid density depends on the relative angle between
the non-Hermitian field ${\bf g}$ and the superflow.
The definition (\ref{eq:sfd}) agrees precisely with the 
more conventional definition \cite{mefish} of $n_s$ based on the dependence
of $E_g$ on periodic phase variations 
(of wavevector ${\bf k}_0$) 
of the wave function:
\begin{eqnarray} 
n_s&=&\frac{m}{\hbar^2}\lim_{k_0\rightarrow 0}
\frac{\partial^2}{\partial k_0^2}
\left[\frac{E_g\left({\bf k}_0\right)}
{{\mathcal V}}\right].
\end{eqnarray}  
It turns out that the superfluid density is directly proportional to
the vortex part of the inverse tilt modulus, ${c_{44}^v}^{-1}$,
in flux line systems:
\begin{equation}
{c_{44}^v}^{-1}=\frac{n_s}{n^2\tilde\epsilon_1},
\label{eq:tiltm}
\end{equation} 
where $n$ is the density of flux lines.\cite{lark}
In three-dimensional superconductors, $n_s$ and $n$ represent 
{\it areal} densities.

Both ${\bf J}_I$ and $n_s$ are zero in the Bose glass
phase and nonzero in the superfluid (or flux liquid) phase.
By computing these quantities for various parameter values,
we are able to construct a phase diagram associated with 
the Bose-glass to flux-liquid transition.
When combined with the results on the Hartree and Bogoliubov spectra
and associated wave functions, these calculations provide 
a physical picture for the delocalized phase  
and the delocalization transition substantially  
different from the noninteracing case, as discussed below.
 
Let us again consider the one-dimensional problem where there is a mobility
edge separating low-energy localized states from high-energy 
extended ones.
As $H_\parallel$ (corresponding to the boson chemical potential) is increased
from the Bose glass side,
the localized states are filled in order of increasing energy. 
At the same time, the interaction effect, which is proportional
to the density and screens the random 
potential, becomes stronger. Since the screening effect is most
effective for the many-body ground state, this 
ground state becomes 
extended one at a critical value of $H_\parallel$, {\it well before} 
the chemical potential hits the mobility edge
in the single particle picture, and the system becomes 
a tilted flux liquid. As $H_\parallel$ is increased further, more
low energy states are transformed to extended ones
and eventually all low-energy states will become extended.
These arguments suggest that, close to the Bose glass
phase boundary, the tilted flux liquid is characterized by 
a low-energy band of Bogoliubov excitations delocalized by hopping,
a band of 
{\it localized excitations}, and then 
a band of more conventional delocalized states.
Its spectrum thus contains {\it two}
mobility edges.

The outline of the paper is as follows.
In Sec.~\ref{sec:nhhb}, we develop the basic formalism using the Hartree
and Bogoliubov approximations.
In Sec.~\ref{sec:spec}, we present the numerical results in one dimension 
on the energy spectra and associated 
eigenfunctions both in a random potential and in the presence of
a single impurity. The nature of the eigenfunctions is studied by computing
the participation ratios and  
the winding numbers. 
In Sec.~\ref{sec:tilt}, we present the numerical results on the
vortex tilt angle and the tilt modulus of ($1+1$)-dimensional flux
line systems both in a random potential and in the presence of a single 
impurity. We also construct the phase diagram associated with the
flux-liquid to Bose-glass (or Bose-insulator) transition.
In Sec.~\ref{sec:conc}, we conclude the paper with some remarks.
In Appendix~\ref{sec:bog}, we develop an analytical theory of both
continuum and lattice non-Hermitian models of a Bose gas in 
a weak random potential using
the Bogoliubov approximation.

\section{Non-Hermitian Hartree-Bogoliubov theory}  
\label{sec:nhhb} 
 
\subsection{Model} 
\label{sec:model} 

Our main interest is in understanding the physics
of flux lines in high-temperature superconductors
with columnar defects. 
Therefore it is useful to clarify the connection 
between the boson Hubbard model (\ref{eq:bhhlb})
and the model free energy $F$ 
for $N$ flux lines 
in a sample of thickness $L_z$ in the $z$ direction 
(perpendicular to the ${\rm CuO}_2$ planes) 
in the presence of $M$ columnar defects aligned in  
the $z$ direction, but randomly distributed 
in the $xy$ plane:
\begin{eqnarray} 
&&F=\frac{\tilde\epsilon_1}{2}\sum_{i=1}^N 
\int_0^{L_z}\left\vert\frac{d{\bf r}_i(z)}{dz} 
\right\vert^2dz 
\nonumber\\ 
&&~~~~~+\frac{1}{2}\sum_{i\ne j}\int_0^{L_z} 
{\mathcal U}\left(\left|{\bf r}_i(z)-{\bf r}_j(z)\right|\right)dz 
\nonumber\\ 
&&~~~~~+\sum_{i}\int_0^{L_z}V_D\left[{\bf r}_i(z)\right]dz 
\nonumber\\ 
&&~~~~~-\frac{\phi_0{\bf H}_\perp}{4\pi}\cdot\sum_{i} 
\int_0^{L_z}\frac{d{\bf r}_i(z)}{dz}dz, 
\nonumber\\
&&V_D\left[{\bf r}_i(z)\right] 
=\sum_{k=1}^MV_1
\left[\left|{\bf r}_i(z)-{\bf R}_k\right|\right],
\label{eq:gl}  
\end{eqnarray} 
where ${\bf r}_i$ and ${\bf R}_k$ are two-dimensional
vectors in the $xy$ plane.
This free energy is a good approximation if
the mass anisotropy $m_z/m_\perp$ in the underlying Ginzburg-Landau 
model is sufficiently large, as is the case
in cuprate superconductors.\cite{vino}
${\mathcal U}(r)$ is the repulsive interaction 
potential between flux lines, 
which can be taken to be local in $z$, 
while $V_1$ is the (attractive) interaction between a flux line 
and a columnar defect. 
The effective line tension
\begin{equation} 
\tilde\epsilon_1=\frac{m_\perp}{m_z}\left(\frac{\phi_0}{4\pi\lambda}
\right)^2\ln \kappa,
\end{equation} 
where $\kappa=\lambda/\xi$ is the ratio of the London penetration
depth $\lambda$ and the coherence length $\xi$,
arises from the tilt energy 
of the lines. 

The canonical partition
function for a system of $N$ lines is 
given by the path integral  
\begin{equation}
Z=\prod_{i=1}^N\int{\mathcal D}\left[{\bf r}_i(z)\right]
e^{-\beta F\left[\left\{{\bf r}_i(z)\right\}\right]},
\end{equation}
which can be rewritten in terms of the imaginary-time
evolution operator
$e^{-L_\tau {\mathcal H}/\hbar}$ as
\begin{equation}
Z=\left\langle\psi^f\left|e^{-L_\tau {\mathcal H}/\hbar}
\right|\psi^i\right\rangle,
\end{equation} 
where the bra and ket vectors are the initial and final states,
respectively.
The lattice version of the quantum Hamiltonian appearing 
in the above equation is precisely the boson Hubbard Hamiltonian
(\ref{eq:bhhlb}).
The parameters of this model, expressed in the original  
flux line language, are summarized in Table~\ref{table1}.
 
In the main part of this paper, we will concentrate 
on the ($1+1$)-dimensional  
case, though the theory developed here can also be applied to ($2+1$) 
dimensions. For convenience, 
we rewrite the Hamiltonian ({\ref{eq:bhhlb}) as 
\begin{eqnarray} 
{\mathcal H}-\mu N&=&-\frac{t}{2}\sum_{i}\left( 
e^ha_{i+1}^\dagger a_i+e^{-h}a_i^\dagger a_{i+1}\right) 
\nonumber\\ 
&&+\sum_i\left(V_i-\mu\right) a_i^\dagger a_i 
+\frac{U}{2}\sum_ia_i^\dagger a_i^\dagger a_ia_i, 
\label{eq:bhh} 
\end{eqnarray} 
where $h=ga/\hbar$.
We assume that the total number of lattice sites is $L$. The 
density is then equal to $n\equiv N/L$.

\subsection{Hartree approximation} 
\label{sec:hart} 

In general, {\it right} eigenfunctions of 
a non-Hermitian operator are not simple complex
conjugates of the corresponding {\it left} eigenfunctions.  
Let us consider the right Hartree state 
\begin{equation} 
\psi^R\left(r_1,r_2,\cdots,r_N\right) 
=\phi_0^R\left(r_1\right)\phi_0^R\left(r_2\right) 
\cdots\phi_0^R\left(r_N\right),
\label{eq:psir} 
\end{equation} 
where all bosons are condensed into the  
same state described by the 
normalized single-particle right eigenfunction 
$\phi_0^R(r_i)$, as a variational ground state for (\ref{eq:bhh}). 
The corresponding left Hartree state is 
\begin{equation} 
\psi^L\left(r_1,r_2,\cdots,r_N\right) 
=\phi_0^L\left(r_1\right)\phi_0^L\left(r_2\right) 
\cdots\phi_0^L\left(r_N\right).
\label{eq:psil} 
\end{equation}
The many-particle states (\ref{eq:psir}) and (\ref{eq:psil})
are explicitly symmetric under particle exchange.
 
The expectation value of the Hamiltonian (\ref{eq:bhh}) 
between the left and right Hartree states is  
\begin{eqnarray} 
&&\langle\psi^L \vert \left({\mathcal H}-\mu N\right)\vert\psi^R\rangle= 
\nonumber\\&& 
~~~~-\frac{tN}{2}\sum_i\left[e^h\phi_0^L(i+1)\phi_0^R(i)+ 
e^{-h}\phi_0^L(i)\phi_0^R(i+1)\right] 
\nonumber\\ 
&&~~~~+N\sum_i\left(V_i-\mu\right) \phi_0^L(i)\phi_0^R(i) 
\nonumber\\ 
&&~~~~+\frac{UN^2}{2}\sum_i\left[\phi_0^L(i)\phi_0^R(i) 
\right]^2. 
\label{eq:expv} 
\end{eqnarray} 
In order to minimize Eq.~(\ref{eq:expv})  
with respect to the single-particle 
state $\phi_0^L(i)$, we need to solve the discrete  
nonlinear Schr\"odinger 
equation: 
\begin{eqnarray} 
&&-\frac{t}{2}\left[e^h\phi_\lambda^R(i-1)+e^{-h}\phi_\lambda^R(i+1) 
\right]+W_i\phi_\lambda^R(i) 
\nonumber\\&&~~~~~~~~~~~~~~~~~~~~~~~~~~~~~~~~~~~~~~~= 
(\mu+\epsilon_\lambda)\phi_\lambda^R(i), 
\label{eq:sch1} 
\end{eqnarray} 
where the effective single-particle potential 
$W_i$ is defined by 
\begin{equation} 
W_i =V_i+UN\phi_0^L(i)\phi_0^R(i). 
\label{eq:ep} 
\end{equation} 
Similarly, we find that the wave function $\phi_\lambda^L(i)$ 
satisfies 
\begin{eqnarray} 
&&-\frac{t}{2}\left[e^{-h}\phi_\lambda^L(i-1)+e^{h}\phi_\lambda^L(i+1) 
\right]+W_i\phi_\lambda^L(i) 
\nonumber\\&&~~~~~~~~~~~~~~~~~~~~~~~~~~~~~~~~~~~~~~~= 
(\mu+\epsilon_\lambda)\phi_\lambda^L(i). 
\label{eq:sch2} 
\end{eqnarray} 
$\phi_0^R(i)$ and $\phi_0^L(i)$ are the ground state wave 
functions of Eqs.~(\ref{eq:sch1}) and (\ref{eq:sch2}).
We assume that the wave functions satisfy the periodic 
boundary conditions appropriate to Fig.~\ref{fig:pbc} 
\begin{eqnarray} 
\phi_\lambda^R(L)&=&\phi_\lambda^R(0),~~~ 
\phi_\lambda^R(L+1)=\phi_\lambda^R(1), 
\nonumber\\ 
\phi_\lambda^L(L)&=&\phi_\lambda^L(0),~~~ 
\phi_\lambda^L(L+1)=\phi_\lambda^L(1). 
\label{eq:pbc} 
\end{eqnarray}

Due to the simplicity of the Hamiltonian and the boundary
conditions, the eigenfunctions $\phi_\lambda^{R,L}(i)$ 
and 
the eigenvalues $\epsilon_\lambda$ satisfy simple symmetry
relationships\cite{hat}
\begin{eqnarray}
\phi_\lambda^L(i;h)&=&\left[\phi_\lambda^R(i;-h)\right]^*,
\nonumber\\
\epsilon_\lambda(h)&=&\left[\epsilon_\lambda(-h)\right]^*.
\end{eqnarray} 
The chemical potential  
$\mu$ is chosen so that the (real-valued) eigenvalue $\epsilon_0$ 
is equal to zero. This condition ensures that the Bogoliubov quasiparticle 
spectrum has no gap, in conformity with general theorems.\cite{pines} 
From Eq.~(\ref{eq:sch1}) or (\ref{eq:sch2}), we then find 
\begin{eqnarray} 
\mu&=&-\frac{t}{2}\sum_i\left[e^h\phi_0^L(i+1)\phi_0^R(i) 
+e^{-h}\phi_0^L(i)\phi_0^R(i+1)\right]\nonumber\\ 
&&+\sum_i V_i\phi_0^L(i)\phi_0^R(i) 
+UN\sum_i\left[\phi_0^L(i) 
\phi_0^R(i)\right]^2. 
\label{eq:cp} 
\end{eqnarray} 
The ground state $\phi_0^R(i)$ ($\phi_0^L(i)$) 
and the excited states $\{\phi_{\lambda\ne 0}^R(i)\}$ 
($\{\phi_{\lambda\ne 0}^L(i)\}$) form an orthonormal basis for  
single-particle states. In Sec.~\ref{sec:spec}, we will 
solve Eqs.~(\ref{eq:sch1}-\ref{eq:sch2}) 
numerically in a self-consistent manner.

\subsection{Bogoliubov approximation} 
\label{sec:bogol} 

In the Bogoliubov approximation, one assumes that the 
ground state (or the condensate) is {\it macroscopically} occupied
with occupation number $N_0$.
This assumption is not true in 
(or sufficiently close to) the Bose glass phase in any dimension.
Even though the superfluid to Bose-glass transition is possible
in ($1+1$) dimensions, phase fluctuations will actually lead to
algebraic decay instead of long-range order in the
boson order parameter. We illustrate this behavior in Fig.~\ref{fig:x},
where we plot the occupation number of the Hartree eigenstates,
$N_\lambda=\langle b_\lambda^L b_\lambda^R\rangle$ (see Eq.~(\ref{eq:ocn})), 
measured in the 
Bogoliubov ground state against the winding number $N_w$ of the 
corresponding Hartree eigenfunctions. We find that the occupation number
rises sharply as the ground state with $N_w=0$ is approached. Though
the ground state is not macroscopically occupied, it still plays a 
dominant role in long-wavelength physics
and the Bogoliubov approximation can be used 
to calculate properly quantities such as
the imaginary current and the superfluid density.\cite{popov}
Our theory is basically a mean-field approach and is expected
to be valid away from the critical region.

From now on, sums over $\lambda$ ($=1,2,\cdots,L-1$) 
will always exclude the condensate. 
We expand the boson field operator $a_i$ in the complete set of  
right annihilation operators $b_0^R$ and $\left\{b_\lambda^R\right\}$: 
\begin{equation} 
a_i=\phi_0^R(i)b_0^R+\sum_{\lambda}\phi_\lambda^R(i)b_\lambda^R. 
\label{eq:a1} 
\end{equation} 
Similarly, we expand $a_i^\dagger$ in terms of the left 
annihilation operators $b_0^L$ and $\left\{b_\lambda^L\right\}$: 
\begin{equation} 
a_i^\dagger=\phi_0^L(i)b_0^L+\sum_\lambda\phi_\lambda^L(i)b_\lambda^L. 
\label{eq:a2} 
\end{equation}  
In the Bogoliubov approximation, we replace the operators $b_0^L$ 
and $b_0^R$ by $\sqrt{N_0}$, where $N_0$ is the
number of bosons in the condensate. The total number of bosons 
is the sum of those 
in the condensate and those not in the condensate: 
\begin{equation} 
N=N_0+\sum_\lambda b_\lambda^Lb_\lambda^R. 
\label{eq:ocn}
\end{equation} 
Expanding to first order in the depletion of the condensate, we get 
\begin{equation} 
b_0^L=b_0^R=\sqrt{N_0}=\sqrt{N}-\frac{1}{2\sqrt{N}}\sum_\lambda 
b_\lambda^Lb_\lambda^R+\cdots. 
\label{eq:expn} 
\end{equation} 
Substituting Eqs.~(\ref{eq:a1}), (\ref{eq:a2}) and (\ref{eq:expn}) 
into Eq.~(\ref{eq:bhh}) and keeping all terms second order in  
$b_\lambda^L$ and $b_\lambda^R$, we obtain a quadratic Bogoliubov-type 
Hamiltonian 
\begin{eqnarray} 
&&{\mathcal H}_B-\mu N=-\frac{UN^2}{2}\sum_i\left[\phi_0^L(i) 
\phi_0^R(i)\right]^2 
\nonumber\\ 
&&~~+\sum_{\lambda\lambda^\prime}\left(A_{\lambda\lambda^\prime} 
b_\lambda^Lb_{\lambda^\prime}^R+\frac{1}{2}B_{\lambda\lambda^\prime} 
b_\lambda^Lb_{\lambda^\prime}^L 
+\frac{1}{2}C_{\lambda\lambda^\prime}b_\lambda^Rb_{\lambda^\prime}^R 
\right), 
\label{eq:bogh0} 
\end{eqnarray} 
where the matrices $A$, $B$ and $C$ are defined by 
\begin{eqnarray} 
A_{\lambda\lambda^\prime}&=&\epsilon_\lambda\delta_{\lambda\lambda^\prime} 
+UN\sum_i\phi_0^L(i)\phi_0^R(i)\phi_\lambda^L(i) 
\phi_{\lambda^\prime}^R(i), 
\nonumber\\ 
B_{\lambda\lambda^\prime}&=&UN\sum_i\left[\phi_0^R(i)\right]^2 
\phi_\lambda^L(i) 
\phi_{\lambda^\prime}^L(i), 
\nonumber\\ 
C_{\lambda\lambda^\prime}&=&UN\sum_i\left[\phi_0^L(i)\right]^2 
\phi_\lambda^R(i) 
\phi_{\lambda^\prime}^R(i). 
\end{eqnarray} 
The matrices $B$ and $C$ describe the processes 
(including effects due to the random potential!) where a pair
of particles are scattered out of and into the condensate respectively,
while the second term of the matrix A describes single-particle
scattering by the condensate.

We diagonalize the Bogoliubov Hamiltonian (\ref{eq:bogh0}) 
by introducing a set of quasiparticle creation and annihilation 
operators $\gamma_\mu^\dagger$ and $\gamma_\mu$  
($\mu = 1, 2, \cdots, L-1$),   
defined by the {\it non-unitary} canonical 
transformation  
\begin{eqnarray} 
\gamma_\mu^\dagger&=&\sum_\lambda\left(u_{\mu\lambda}b_\lambda^L 
-v_{\mu\lambda}b_\lambda^R\right),
\nonumber\\
\gamma_\mu&=&\sum_\lambda\left({\overline u}_{\mu\lambda}b_\lambda^R 
-{\overline v}_{\mu\lambda}b_\lambda^L\right),  
\end{eqnarray}
where the coefficients 
${\overline u}_{\mu\lambda}$ and ${\overline v}_{\mu\lambda}$ 
are defined by
\begin{equation}
{\overline u}_{\mu\lambda}(h)=\left[u_{\mu\lambda}(-h)\right]^*,
~~{\overline v}_{\mu\lambda}(h)=\left[v_{\mu\lambda}(-h)\right]^*.
\end{equation}
The quasiparticle operators satisfy 
\begin{eqnarray} 
\left[{\mathcal H}_B,\gamma_\mu^\dagger\right]&=&\omega_\mu\gamma_\mu^\dagger, 
\nonumber\\ 
\left[{\mathcal H}_B,\gamma_\mu\right]&=&-\omega_\mu\gamma_\mu, 
\end{eqnarray} 
where $\omega_\mu$ is the (complex-valued) quasiparticle excitation 
energy.  
The transformation coefficients  
$u_{\mu\lambda}$ and $v_{\mu\lambda}$ turn out to 
obey the eigenvalue equations 
\begin{eqnarray} 
\sum_{\lambda^\prime}\left(A_{\lambda\lambda^\prime} 
u_{\mu\lambda^\prime}+B_{\lambda\lambda^\prime} 
v_{\mu\lambda^\prime}\right)&=&\omega_\mu u_{\mu\lambda}, 
\nonumber\\ 
\sum_{\lambda^\prime}\left(C_{\lambda\lambda^\prime} 
u_{\mu\lambda^\prime}+A_{\lambda^\prime\lambda} 
v_{\mu\lambda^\prime}\right)&=&-\omega_\mu v_{\mu\lambda} 
\label{eq:bspec} 
\end{eqnarray} 
and the orthonormality conditions
\begin{equation}
\sum_{\lambda}\left({\overline u}_{\mu\lambda}u_{\mu^\prime\lambda}-
{\overline v}_{\mu\lambda}v_{\mu^\prime\lambda}\right)
=\delta_{\mu\mu^\prime}.
\label{eq:bnorm}
\end{equation}
In Sec.~\ref{sec:spec},
we solve these equations numerically to obtain the coefficients 
$u_{\mu\lambda}$ and $v_{\mu\lambda}$ and the eigenvalues $\omega_\mu$.
Finally, using Eq.~(\ref{eq:cp}), we find that  
the diagonalized Bogoliubov Hamiltonian has the form 
\begin{eqnarray} 
{\mathcal H}_B&=&E_g 
+\sum_\mu\omega_\mu\gamma_\mu^\dagger\gamma_\mu, 
\nonumber\\ 
\frac{E_g}{N}&=&-\frac{t}{2}\sum_i\left[e^h\phi_0^L(i+1)\phi_0^R(i) 
+e^{-h}\phi_0^L(i)\phi_0^R(i+1)\right]\nonumber\\ 
&&+\sum_i V_i\phi_0^L(i)\phi_0^R(i)
+\frac{Un}{2}L\sum_i\left[\phi_0^L(i) 
\phi_0^R(i)\right]^2
\nonumber\\ 
&&+\frac{1}{2n}\frac{1}{L}\sum_\mu\left(\omega_\mu-A_{\mu\mu}\right), 
\label{eq:hgse} 
\end{eqnarray} 
where $E_g$ is the ground state energy.

\section{Spectra and wave functions} 
\label{sec:spec} 
  
\subsection{Ground state wave function and screening}

The first step of our numerical calculation is to 
solve Eqs.~(\ref{eq:sch1}-\ref{eq:sch2}) 
with the periodic boundary condition (\ref{eq:pbc}).
In all of our calculations, we will fix the hopping
integral $t$ to a constant value.
Then there are three natural dimensionless parameters  
in Eqs.~(\ref{eq:sch1}-\ref{eq:sch2}), that is,   
the non-Hermitian field $h$, the  
interaction parameter $Un/t$ and 
the disorder parameter $\Delta/t$.  
It turns out that the effect of disorder is negligible if 
$h$ or $Un/t$ is sufficiently large, or  
of course if $\Delta/t$ is sufficiently small. 
In these cases, 
we begin with (normalized) trial condensate wave functions 
$\phi_0^L(i)=\phi_0^R(i)=1/\sqrt{L}$, which is uniform in space. 
Then we solve the nonlinear Schr\"odinger equations with  
the screened random potential $W_i=V_i+Un$ numerically to find 
new left and right single-particle eigenstates. We obtain 
updated trial condensate wavefunctions by mixing the initial guess
with the ground state eigenstate of the screened potential, using
the so-called Broyden mixing method widely used in
self-consistent computations of nonlinear equations.\cite{broy}   
This procedure is repeated until a convergence is achieved. 
 
In the presence of disorder, the ground state wave function is nonuniform 
in space. We use the solution obtained for a large $h$ or $Un/t$, or 
for a small $\Delta/t$ as trial wave functions for the cases with 
a smaller $h$ or $Un/t$, 
or with a larger $\Delta/t$. The convergence is very slow if  
the system size is large and the effective disorder is strong. 
For all numerical results presented here, the lattice size is limited 
to $L\le 200$.     
   
In Fig.~\ref{fig:cd},
we plot the local condensate density $n_0(i)=\phi_0^L(i)\phi_0^R(i)$
in a single realization of the random potential
for $h=0.1$, $Un/t=0.5$, $\Delta/t=1$ and $L=60$. We notice that the 
condensate is indeed highly nonuniform. In fact, the local condensate 
density is large (small) at the local minima (maxima)  
of the bare random potential $V_i$
as shown in the inset of Fig.~\ref{fig:cd}, so that the screened random  
potential is smoother than $V_i$. In Fig.~\ref{fig:srp}, we compare the 
screened random potential for the same parameter values 
as in Fig.~\ref{fig:cd} with the corresponding bare random potential. 
The screening is more effective for larger values of $Un/t$.  
We have verified that the screening does not produce long-range  
correlations in the random potential, 
as illustrated in Fig.~\ref{fig:cor}. 
 
\subsection{Hartree and Bogoliubov spectra} 

After calculating the self-consistent eigenvalues
and eigenfunctions using the Hartree approximation, 
we use them to solve the Bogoliubov eigenvalue equations
(\ref{eq:bspec}) and (\ref{eq:bnorm}). The eigenvalues
$\omega_\mu$ obtained from this calculation give the Bogoliubov
quasiparticle spectrum.
 
In Fig.~\ref{fig:spec1}(a), we show the Hartree spectra
of a 200-site lattice
in a single disorder configuration  
for $Un/t=0.5$, $\Delta/t=1$  
and $h=0.8, 0.6, 0.3$.
For these parameter values, the system is deep inside
the superfluid region. 
When the non-Hermitian field is strong, 
the spectrum is close to that of a clean system, with  
no real eigenvalues except at the very bottom and top of the band.
All states are extended in this case. 
As the non-Hermitian field $h$ is reduced, some of the  
eigenvalues near the top of the band become real at some  
critical value of $h$. This band of real-valued eigenvalues
corresponding to localized states 
grows continuously as $h$ is further reduced.
We note that there is a mobility edge separating low-energy
extended states from high-energy localized states.
  
In Fig.~\ref{fig:spec1}(b), 
we show the {\it Bogoliubov} excitation 
spectra for the same parameter values 
as in Fig.~\ref{fig:spec1}(a). 
Similarly to the Hartree spectra, there is 
a low-energy complex bubble corresponding to a band of 
extended states. For small values of $h$, 
a band of (localized) states with real eigenvalues appears 
at the top of the bubble. The Bogoliubov spectrum shows 
a linear behavior near the ground state,
similar to results for non-Hermitian ``superfluids''
without disorder.\cite{lehr} This linear vanishing of
Im~$\omega_\mu$ with Re~$\omega_\mu$  
is due to the collective phonon excitations 
characteristic of the superfluid phase. We notice that the
effect of disorder is weaker in the Bogoliubov spectra
than in the Hartree spectra in the sense that 
the critical value of $h$
where localized states begin to appear at the top of the band
is smaller.   

Next, we consider the change of the Hartree and Bogoliubov spectra
as we cross the Bose glass-superfluid phase boundary 
by increasing the value of $h$, starting from
the Bose glass side. When $h$ is zero, all eigenvalues are real
and, in a sufficiently disordered system, low-lying eigenstates
including the ground state are localized. As $h$ is increased,
a bubble of extended states 
with complex eigenvalues appears in the
center of the band, but the low-energy states with real eigenvalues
remain localized. As $h$ crosses the phase boundary, the ground
state becomes extended and the system is transformed to a superfluid. 
Within the mean-field theory
framework, the {\it condensate} is formed, even though it exhibits
a slow algebraic decay at $T=0$ in ($1+1$) dimensions.
As $h$ is increased further, a few lowest-lying states become extended
and form a small bubble in the complex plane. Note that there 
is a small band of {\it localized excited states} sandwiched between two
bands of extended states in this parameter regime!
The entire spectrum thus contains {\it three} mobility edges. 
In the continuum limit where lattice effects can be neglected,
we expect the two low-energy mobility edges survive, while the top one 
disappears. For larger values of $h$, two
bubbles grow in size and eventually merge. This
``double bubble'' scenario is illustrated in Fig.~\ref{fig:spec2}
for a single disorder configuration and for $Un/t=0.05$, $\Delta/t=1$,
$h=0.3, 0.44, 0.5$ and $L=60$. Again, the effect of disorder
is weaker in the Bogoliubov spectra than in the Hartree spectra,
and the characteristic linear behavior in the Bogoliubov spectrum
is observed in the small bubble including the ground state. 
We have checked that this ``double bubble'' scenario
is quite robust and occurs in any realization of the random potential.

\subsection{Participation ratio and winding number} 
 
The nature of the Hartree eigenstates can be studied by  
calculating the participation ratio\cite{rokh} 
\begin{equation} 
P_\lambda=\frac{1} 
{\sum_i\big\vert\phi_\lambda^L(i)\phi_\lambda^R(i) 
\big\vert^2} 
\end{equation} 
of state $\lambda$. 
This quantity is a measure of the number of sites where 
the wave function is not negligible. In case of plane
waves, $P_\lambda$ is equal to the total number of sites
and diverges in the thermodynamic limit. For more general 
extended states, a weaker condition 
\begin{equation}
\lim_{L\rightarrow\infty}\ln P_\lambda\propto\ln L
\end{equation}
is applied. For a localized state, $P_\lambda$ is 
proportional to the localization volume and approaches
a constant in the thermodynamic limit.

In Fig.~\ref{fig:pr}, 
we plot the participation ratios for $Un/t=0.5$, 
$\Delta/t=1$, $h=0.8, 0.3$ and $L=100, 200$. 
We find that, for $h=0.8$, the participation ratios for 
all states scale with the size of the system. This can be  
considered as evidence that all eigenstates in this  
case are extended. 
When $h=0.3$, the participation ratios for the states 
having real eigenvalues at the top of the band do not scale  
with the size of the system, 
while those for the states with complex eigenvalues 
and the ground state do.   
We conclude that the states near the top of the band  
with real eigenvalues are localized and there exists  
a mobility edge separating them from 
low-energy extended states.  
 
A better criterion for determining the nature of Hartree  
eigenstates in one dimension is provided by the winding number  
of the Hartree eigenfunction.\cite{shn}
This winding number is the generalization for non-Hermitian 
disordered systems of states labelled by wavevectors
in translationally invariant problems.
When the eigenvalue $\epsilon_\lambda$ is complex, the eigenfunction
$\phi_\lambda^{R,L}(i)$ rotates an integer times around the origin
on the complex plane, as the spatial dimension (which is a circle)
is traversed once. This integer rotation or winding number 
$N_w$ is well-defined
even in strongly-disordered cases. It has been argued that the states
with nonzero winding numbers are always extended.\cite{shn} 
  
In Fig.~\ref{fig:wn}(a), we compare the Hartree spectrum and 
the winding numbers for $h=0.3$, $Un/t=0.5$, $\Delta/t=1$ and $L=200$.
In Fig.~\ref{fig:wn}(b), we show the curves for $h=0.44$, 
$Un/t=0.05$, $\Delta/t=1$ and $L=60$. A bubble of extended states including
the ground state is clearly seen. 
If the eigenvalue is real, the corresponding eigenfunction is 
also real, and the winding number is undefined. In this case, 
we arbitrarily set $N_w=0$. 
 
\subsection{Non-Hermitian vortex physics with a single impurity} 
 
Some features of the Hartree spectrum can be better understood 
by examining the cases with a single impurity.  
In Fig.~\ref{fig:ospec1}(a), we show the spectra in the presence of a  
single attractive impurity with the strength $V_d=-5t$.  
In the noninteracting case, 
there is a single bound state, which is the ground state.\cite{hat} 
As the interaction parameter $Un/t$ increases, this bound 
state moves toward the band of extended states with higher real parts 
of the eigenvalues and, at the same time,  
the highest energy eigenstate becomes 
a bound state. Above some critical value of $Un/t$, the ground state  
merges into the band of extended states and another bound state  
appears at the top of the band.
 
This behavior can be understood from the shape of the  
screened potential $W_i$ shown in Fig.~\ref{fig:ospec1}(b).  
As $Un/t$ increases, the potential well 
of the screened potential becomes shallower and the ground 
state changes from a bound state to an extended one. 
On the other hand, the inverted screened potential, $-W_i$, has two 
potential wells, which can be sufficiently deep to support one or two 
bound states for intermediate values of $Un/t$. This observation explains the  
appearance of {\it hole} bound states at the top of the band.
Extending the present argument to the cases with many (but dilute)
impurities with random strengths, we can qualitatively understand
the Hartree spectra in the presence of a random potential 
shown in Figs.~\ref{fig:spec1} and \ref{fig:spec2}.
 
In Fig.~\ref{fig:ospec2}, we show the spectra and the screened potential
in the presence of a single 
repulsive impurity with the strength $V_d=5t$. 
As $Un/t$ increases, the hole bound state with the highest energy 
moves toward the band of extended states and merges with it at a critical
value of $Un/t$.
This can also be understood from the shape of the screened potential.

\section{Tilt angle, tilt modulus and the Bose-glass to flux-liquid 
transition} 
\label{sec:tilt} 
 
After obtaining the Hartree eigenvalues and eigenfunctions 
and the Bogoliubov eigenvalues, we use them to compute the ground
state energy per particle, $E_g/N$, given in Eq.~(\ref{eq:hgse}).
We note that the last term of Eq.~(\ref{eq:hgse}) has an explicit
dependence on the density $n$, while the first three terms do not.
This suggests that the density is another independent parameter
in addition to $Un/t$, $\Delta/t$ and $h$. For all parameter values
used in this work, the last term of Eq.~(\ref{eq:hgse}) is numerically
much smaller than the other terms, unless the density is very small.
In the rest of our calculations, we will fix $n=0.6$.   

At the final stage of our calculations, we take numerical derivatives of 
the ground state energy with respect to the non-Hermitian field $h$
and calculate the vortex tilt angle, $\theta_v$, and the vortex 
component of the inverse tilt modulus, ${c_{44}^v}^{-1}$, 
using Eqs.~(\ref{eq:tilta}),
(\ref{eq:sfd}) and (\ref{eq:tiltm}). 
In Fig.~\ref{fig:tranh}, we plot $\theta_v$ and ${c_{44}^v}^{-1}$
as functions of $h$, which is proportional
to the transverse external field $H_\perp$, for different
values of $Un/t$. It is clearly
seen that there is a transition from the Bose glass phase with
$\theta_v=0$ and $c_{44}^v=\infty$ to the flux liquid phase with
both quantities finite, as $h$ is increased through a critical value $h_c$.
The transition is broader when the interaction is stronger.
It has been predicted in Ref.~24 that, close to the phase boundary,
${c_{44}^v}^{-1}\sim (h-h_c)^{1/2}$. Our calculations are not precise
enough to give critical exponents.

In Fig.~\ref{fig:phase}, we show an approximate phase diagram in 
the $U$-$h$ plane 
constructed using the calculations done on a single-size lattice
($L=30$) 
for a single realization of the random 
potential with $\Delta/t=1$.    
We chose the first inflection point of the ${c_{44}^v}^{-1}$ curve
plotted against $h$ as the rough phase boundary. 
As the interaction strength increases,
the system becomes effectively less disordered, 
and therefore the critical value of $h$
required to drive the system to the flux liquid phase becomes smaller.

In Fig.~\ref{fig:fluc}, we plot the curves for $\theta_v$ and 
${c_{44}^v}^{-1}$ versus $h$ 
obtained for five different realizations of the random 
potential. Due to the small system size ($L=30$) 
used in the calculations, these quantities show sizable
fluctuation effects. The relative fluctuation of the vortex tilt angle 
is seen to be smaller
than that of the inverse tilt modulus.

The Bose-glass to flux-liquid transition can also be driven by changing
parameters such as the disorder strength $\Delta/t$ and 
the density $n$. In
Fig.~\ref{fig:trand}, we show $\theta_v$ and ${c_{44}^v}^{-1}$
as functions of $\Delta/t$. A transition is observed at a critical strength
of $\Delta/t$. When the non-Hermitian field $h$ is bigger, the transition 
occurs at a higher value of $\Delta/t$. In this figure, we also
compare the numerical results with the analytical formulae in
the weak disorder limit derived in Appendix A. Since the variance of the
random potential, $\langle V_i^2\rangle-\langle V_i\rangle^2$,
for the particular disorder configuration used in this calculation
is approximately 0.419, we relate the disorder parameter in the 
analytical formulae, $\Gamma$, to the parameter $\Delta$ by
$\Gamma\approx 0.419\Delta^2$. 

We have also computed the tilt angle and the tilt modulus for the cases
with a single attractive impurity. When both the interaction 
and the transverse external field is very weak,
all flux lines can be pinned by one impurity and the system is an 
insulator. As the interaction strength is increased however, 
the screening effect
causes the lowest-energy bound state to change into 
an extended one and the system
becomes a tilted flux liquid. This is illustrated in Fig.~\ref{fig:trano}
and the corresponding phase diagram is shown in Fig.~\ref{fig:phaseo}.

\section{Conclusion} 
\label{sec:conc}

In the present paper, we have developed a numerical method of studying
the strongly-disordered boson Hubbrad model in the presence
of a constant imaginary vector potential and a random scalar potential
and used it to understand the physics of vortices in high-temperature
superconductors with columnar defects. 
We have found that the interaction causes the screening of a random 
potential, which is most effective for the ground and low-lying 
excited states. For sufficiently large values of 
the effective interaction parameter or 
the non-Hermitian field, the nature of these states is changed 
from localized 
to extended. All physical properties are strongly influenced
by these phenomena. In this paper, we presented a detailed study
of the vortex tilt angle and the tilt modulus in ($1+1$)
dimensions.

\acknowledgments 
K. Kim is deeply grateful to Jack Lidmar for numerous enlightening 
discussions.
This work has been supported by the NSF through 
Grant No. DMR97-14725 and through the Harvard MRSEC via Grant No.
DMR98-09363 and  
by the Korea Science and Engineering Foundation 
through Grant No. 1999-2-11400-005-5.

\appendix 
 
\section{Non-Hermitian Bogoliubov theory  
of a dilute Bose gas in a weak 
random potential} 
\label{sec:bog}
 
In the {\it weak} disorder limit, it is possible to obtain analytical 
results for various physical quantities using the Bogoliubov 
approximation. In this Appendix, we describe the results for both 
continuum and lattice models of a dilute Bose gas in a constant 
imaginary vector potential and a weak random potential. In the 
section on the continuum model, we define $\mathcal V$ as the volume 
and $n$ as the number of particles per volume.  
In the section on the lattice model,  
$\mathcal V$ represents the number of lattice sites 
and $n$ is the number of particles per site. 
Our conclusions serve as a check on results obtained via
coherent state path integrals in Ref.~7. The results 
for non-Hermitian lattice models with disorder are new. 
 
\subsection{Continuum model} 
 
We consider a dilute Bose gas of volume $\mathcal V$ in a weak random 
potential, $V({\bf x})$, and a constant non-Hermitian vector 
potential, ${\bf g}$, in $d$ dimensions. The  
Hamiltonian is given by 
\begin{eqnarray} 
{\mathcal H}&=&\int d^dx~\psi^\dagger 
\left[\frac{1}{2m}\left(\frac{\hbar}{i}\nabla+i{\bf g}\right)^2 
+V({\bf x})\right]\psi\nonumber\\ 
&&+\frac{v_0}{2}\int d^dx~\psi^\dagger\psi^\dagger\psi\psi, 
\label{eq:ham} 
\end{eqnarray} 
where $\psi({\bf x})$ is the boson field operator 
and $v_0$ is the interaction strength.
This model is a continuum version of the lattice model (\ref{eq:bhhlb}). 
Upon substituting 
\begin{eqnarray} 
\psi({\bf x})&=&\frac{1}{\sqrt{{\mathcal V}}} 
\sum_{\bf k}e^{i{\bf k} 
\cdot{\bf x}}a_{\bf k},\nonumber\\ 
V({\bf x})&=&\frac{1}{\sqrt{{\mathcal V}}} 
\sum_{\bf k}e^{i{\bf k}\cdot{\bf x}}V_{\bf k} 
\end{eqnarray} 
into Eq.~(\ref{eq:ham}), we obtain 
\begin{eqnarray} 
{\mathcal H}&=&\sum_{\bf k} 
\frac{\left(\hbar {\bf k}+i{\bf g}\right)^2}{2m} 
a_{\bf k}^\dagger a_{\bf k} 
+\frac{1}{\sqrt{{\mathcal V}}}\sum_{{\bf k},{\bf q}} 
V_{{\bf k}-{\bf q}} a_{\bf k}^\dagger a_{\bf q} 
\nonumber\\ 
&&+\frac{v_0}{2{\mathcal V}}
\sum_{{\bf k},{\bf k}^\prime,{\bf k}^{\prime\prime}} 
a_{\bf k}^\dagger a_{{\bf k}^\prime}^\dagger 
a_{{\bf k}^{\prime\prime}} 
a_{{\bf k}+{\bf k}^\prime-{\bf k}^{\prime\prime}}. 
\label{eq:bham0} 
\end{eqnarray} 
We assume that   
$V({\bf x})$ is a Gaussian random potential, 
characterized by a single parameter $\gamma$: 
\begin{eqnarray} 
\overline{V({\bf x})V({\bf y})}&=&\gamma\delta({\bf x}-{\bf y}), 
\nonumber\\ 
\overline{V_{\bf k}V_{\bf q}}&=&\gamma\delta_{{\bf k},-{\bf q}}, 
\label{eq:rpot} 
\end{eqnarray} 
where $V_{{\bf k}={\bf 0}}=0$ and 
the bar indicates averaging over disorder. 
 
Following Bogoliubov, 
we replace $a_0$ and $a_0^\dagger$ by 
\begin{equation}  
\sqrt{N_0}\approx \sqrt{N}-\frac{1}{2\sqrt{N}}\sum_{{\bf k}\ne {\bf 0}} 
a_{\bf k}^\dagger a_{\bf k} 
\end{equation} 
and 
expand the Hamiltonian in terms of  
$a_{\bf k}$ and $a_{\bf k}^\dagger$ with ${\bf k}\ne {\bf 0}$: 
\begin{eqnarray} 
{\mathcal H}&=&{\mathcal V}\left(-\frac{g^2}{2m} 
n+\frac{1}{2}v_0n^2\right) 
\nonumber\\ 
&&+\sum_{{\bf k}\ne{\bf 0}}\left(\frac{\hbar^2k^2}{2m} 
+v_0n+i\frac{\hbar}{m}{\bf k}\cdot{\bf g} 
\right)a_{\bf k}^\dagger a_{\bf k} 
\nonumber\\ 
&&+\frac{1}{2}v_0n\sum_{{\bf k}\ne{\bf 0}}\left( 
a_{\bf k} a_{-{\bf k}} +a_{\bf k}^\dagger a_{-{\bf k}}^\dagger\right) 
\nonumber\\ 
&&+\sqrt{n}\sum_{{\bf k}\ne{\bf 0}} 
V_{\bf k} \left(a_{\bf k}^\dagger +a_{-{\bf k}}\right) 
+ \cdots, 
\end{eqnarray} 
where $n=N/{\mathcal V}$. 
We ignore the cubic and quartic terms represented by $\cdots$  
and diagonalize the remaining 
terms by the {\it nonunitary} canonical transformation 
\begin{eqnarray} 
a_{\bf k}&=&\frac{c_{\bf k}^R-\alpha_k c_{-{\bf k}}^L} 
{\sqrt{1-\alpha_k^2}}\nonumber\\ 
&&-\sqrt{n}\frac{V_{\bf k}} 
{\omega_{\bf k}\omega_{-{\bf k}}} 
\left(\frac{\hbar^2k^2}{2m}-i\frac{\hbar}{m}{\bf k}\cdot 
{\bf g}\right), 
\nonumber\\ 
a_{\bf k}^\dagger&=&\frac{c_{\bf k}^L-\alpha_k c_{-{\bf k}}^R} 
{\sqrt{1-\alpha_k^2}}\nonumber\\ 
&&-\sqrt{n}\frac{V_{\bf k}^*} 
{\omega_{\bf k}\omega_{-{\bf k}}} 
\left(\frac{\hbar^2k^2}{2m}-i\frac{\hbar}{m}{\bf k}\cdot 
{\bf g}\right), 
\label{eq:can} 
\end{eqnarray} 
where 
\begin{eqnarray} 
\alpha_k&=&\frac{1}{v_0n}\left[\frac{\hbar^2k^2}{2m}+v_0n 
-\sqrt{\frac{\hbar^2k^2}{2m} 
\left(\frac{\hbar^2k^2}{2m}+2v_0n\right)}\right],\nonumber\\ 
\omega_{\bf k}&=&\sqrt{\frac{\hbar^2k^2}{2m} 
\left(\frac{\hbar^2k^2}{2m}+2v_0n\right)}+i\frac{\hbar}{m}{\bf k}\cdot 
{\bf g}. 
\end{eqnarray} 
The left and right annihilation operators 
$c_{\bf k}^L$ and $c_{\bf k}^R$ satisfy bosonic commutation relations 
\begin{equation} 
\left[c_{\bf k}^R,c_{\bf q}^L\right]=\delta_{{\bf k},{\bf q}},~~ 
\left[c_{\bf k}^R,c_{\bf q}^R\right]=0,~~ 
\left[c_{\bf k}^L,c_{\bf q}^L\right]=0. 
\end{equation} 
The diagonalized Bogoliubov Hamiltonian has the simple form 
\begin{eqnarray} 
&&{\mathcal H}_B=E_g 
+\sum_{{\bf k}\ne{\bf 0}}\omega_{\bf k}c_{\bf k}^Lc_{\bf k}^R, 
\nonumber\\ 
&&\frac{E_g}{{\mathcal V}} 
=-\frac{g^2}{2m}n 
+\frac{1}{2}v_0n^2 
\nonumber\\ 
&&~~-\frac{1}{2}\frac{1}{{\mathcal V}}\sum_{{\bf k}\ne{\bf 0}} 
\left[\frac{\hbar^2k^2}{2m}+v_0n 
-\sqrt{\frac{\hbar^2k^2}{2m} 
\left(\frac{\hbar^2k^2}{2m}+2v_0n\right)}\right] 
\nonumber\\ 
&&~~-n\frac{1}{{\mathcal V}} 
\sum_{{\bf k}\ne{\bf 0}}\frac{\vert V_{\bf k}\vert^2} 
{\omega_{\bf k}\omega_{-{\bf k}}}\frac{\hbar^2k^2}{2m}, 
\label{eq:fe} 
\end{eqnarray} 
where $E_g$ is the ground state energy and $\omega_{\bf k}$
is the Bogoliubov quasiparticle spectrum. 
From Eqs.~(\ref{eq:ic}), (\ref{eq:sfd}) and (\ref{eq:fe}), we derive 
the imaginary current,
\begin{eqnarray}
{\bf J}_I&=&\frac{{\bf g}}{m}\Bigg\{1-
\int\frac{d^dk}{(2\pi)^d}~ 
\vert V_{\bf k}\vert^2
\nonumber\\
&&\times\frac{4\cos^2\theta}
{\left[\hbar^2k^2/2m+2v_0n 
+2(g^2/m)\cos^2\theta\right]^2}\Bigg\},
\label{eq:imcc}
\end{eqnarray}
and the superfluid density,
\begin{eqnarray} 
n_s&=&n-4n\int\frac{d^dk}{(2\pi)^d}~ 
\vert V_{\bf k}\vert^2\cos^2\phi\nonumber\\ 
&&\times\Bigg\{\frac{1}{\left[\hbar^2k^2/2m+2v_0n 
+2(g^2/m)\cos^2\theta\right]^2} 
\nonumber\\ 
&&~ 
-\frac{8(g^2/m)\cos^2\theta} 
{\left[\hbar^2k^2/2m+2v_0n 
+2(g^2/m)\cos^2\theta\right]^3}\Bigg\}, 
\label{eq:sd} 
\end{eqnarray} 
where $\theta$ is the angle 
between ${\bf k}$ and ${\bf g}$ and
$\phi$ is the angle between ${\bf k}$ 
and $\delta{\bf g}$. 
In case of the Gaussian random potential,  
the integrals in Eqs.~(\ref{eq:imcc}) and
({\ref{eq:sd}) can be  
performed easily. We summarize the results for  
the disorder-averaged 
imaginary current $\overline{{\bf J}_I}$, 
expressed in terms of the dimensionless 
parameter $\tilde g=g/\sqrt{mv_0n}$, below:
\begin{eqnarray}
\overline{{\bf J}_I}&=&
\frac{{\bf g}}{m}\left[1-\gamma
\frac{1}{2v_0^2n^2}\sqrt{\frac{mv_0n} 
{\hbar^2}}\frac{1}{(1+\tilde g^2)^{3/2}}\right] 
\nonumber\\
&&~~~~~~~~~~~~~~~~~~~~~~~~~~~~~~~~~~~~~~~~\mbox{if 
$d=1$},
\label{eq:ic1}\\ 
\overline{{\bf J}_I}&=&
\frac{{\bf g}}{m}\left[1-\gamma
\frac{m}{2\pi\hbar^2v_0n}\frac{2}{\tilde g^2}\left(
1-\frac{1}{\sqrt{1+\tilde g^2}}\right)
\right]. 
\nonumber\\&& 
~~~~~~~~~~~~~~~~~~~~~~~~~~~~~~~~~~~~~~~~ 
\mbox{if $d=2$}
\end{eqnarray}
The disorder-averaged 
{\it normal} fluid density 
$\overline{n_n}$ ($\equiv n-\overline{n_s}$) is listed below: 
\begin{eqnarray} 
\overline{n_n}&=& 
\gamma \frac{1}{2v_0^2n}\sqrt{\frac{mv_0n} 
{\hbar^2}}\frac{1-2\tilde g^2}{(1+\tilde g^2)^{5/2}} 
~~~\mbox{if 
$d=1$},
\label{eq:nfd1}\\ 
\overline{n_n}&=& 
\gamma\frac{m}{2\pi\hbar^2v_0} 
\frac{2}{\tilde g^2} 
\left[\frac{1+2\tilde g^2}{(1+\tilde g^2)^{3/2}}-1\right] 
\nonumber\\&& 
~~~~~~~~~~~~~~~~~~~~~~~~~~~~~ 
\mbox{if $d=2$, $\delta{\bf g}\parallel{\bf g}$},
\label{eq:nfd2}\\ 
\overline{n_n}&=& 
\gamma\frac{m}{2\pi\hbar^2v_0} 
\frac{2}{\tilde g^2}\left(1-\frac{1}{\sqrt{1+\tilde g^2}}\right) 
\nonumber\\&& 
~~~~~~~~~~~~~~~~~~~~~~~~~~~~~ 
\mbox{if $d=2$, $\delta{\bf g}\perp{\bf g}$}. 
\label{eq:nfd3} 
\end{eqnarray} 
Eqs.~(\ref{eq:nfd1}-\ref{eq:nfd3}) have been derived previously
using a different method.\cite{taub}
 
\subsection{Lattice model} 
 
We consider the $d$-dimensional non-Hermitian Hubbard 
model for lattice bosons, Eq.~(\ref{eq:bhhlb}), in a weak  
random potential $V_{\bf x}$. 
Upon substituting 
\begin{eqnarray} 
a_{\bf x}&=&\frac{1}{\sqrt{\mathcal V}} 
\sum_{\bf k}e^{i{\bf k} 
\cdot{\bf x}}a_{\bf k},\nonumber\\ 
V_{\bf x}&=&\frac{1}{\sqrt{\mathcal V}} 
\sum_{\bf k}e^{i{\bf k}\cdot{\bf x}}V_{\bf k}, 
\end{eqnarray} 
where $\mathcal V$ is the total number of sites 
and $V_{{\bf k}={\bf 0}}=0$, 
into Eq.~(\ref{eq:bhhlb}), we get 
\begin{eqnarray} 
{\mathcal H}&=&\sum_{\bf k}\left[\epsilon_R+i\epsilon_I 
-t\sum_{\nu=1}^d\cosh(g_\nu {a}/\hbar) 
\right]a_{\bf k}^\dagger a_{\bf k} 
\nonumber\\&+&\frac{1}{\sqrt{\mathcal V}} 
\sum_{{\bf k},{\bf q}} 
V_{{\bf k}-{\bf q}} a_{\bf k}^\dagger a_{\bf q} 
\nonumber\\ 
&+&\frac{U}{2\mathcal V}\sum_{{\bf k},{\bf k}^\prime,{\bf k}^{\prime\prime}} 
a_{\bf k}^\dagger a_{{\bf k}^\prime}^\dagger 
a_{{\bf k}^{\prime\prime}} 
a_{{\bf k}+{\bf k}^\prime-{\bf k}^{\prime\prime}}, 
\end{eqnarray} 
where 
\begin{eqnarray} 
\epsilon_R&=&t\sum_{\nu}\cosh(g_\nu {a}/\hbar) 
\left[1-\cos\left( 
k_{\nu}{a}\right)\right], 
\nonumber\\ 
\epsilon_I&=&t\sum_{\nu}\sinh(g_\nu {a}/\hbar)\sin\left( 
k_\nu {a}\right). 
\end{eqnarray} 
Using the Bogoliubov prescription, we obtain an approximate 
Hamiltonian 
\begin{eqnarray} 
{\mathcal H}_B&=&{\mathcal V}\left[-tn\sum_{\nu}\cosh(g_\nu a/\hbar) 
+\frac{1}{2}Un^2\right] 
\nonumber\\ 
&&+\sum_{{\bf k}\ne{\bf 0}}\left(\epsilon_R 
+i\epsilon_I+Un 
\right)a_{\bf k}^\dagger a_{\bf k} 
\nonumber\\ 
&&+\frac{1}{2}Un\sum_{{\bf k}\ne{\bf 0}}\left( 
a_{\bf k} a_{-{\bf k}} +a_{\bf k}^\dagger a_{-{\bf k}}^\dagger\right) 
\nonumber\\&& 
+\sqrt{n}\sum_{{\bf k}\ne{\bf 0}} 
V_{\bf k} \left(a_{\bf k}^\dagger +a_{-{\bf k}}\right), 
\label{eq:bhh2} 
\end{eqnarray} 
where $n$ is the number of particles  
per site.  
We diagonalize this Hamiltonian   
by the canonical transformation 
\begin{eqnarray} 
a_{\bf k}&=&\frac{c_{\bf k}^R-\alpha_{\bf k} c_{-{\bf k}}^L} 
{\sqrt{1-\alpha_{\bf k}^2}}-\sqrt{n}\frac{V_{\bf k}} 
{\omega_{\bf k}\omega_{-{\bf k}}}\left(\epsilon_R-i\epsilon_I 
\right), 
\nonumber\\ 
a_{\bf k}^\dagger&=&\frac{c_{\bf k}^L-\alpha_{\bf k} c_{-{\bf k}}^R} 
{\sqrt{1-\alpha_{\bf k}^2}}-\sqrt{n}\frac{V_{\bf k}^*} 
{\omega_{\bf k}\omega_{-{\bf k}}}\left(\epsilon_R-i\epsilon_I 
\right), 
\label{eq:lcan} 
\end{eqnarray} 
where 
\begin{eqnarray} 
\alpha_{\bf k}&=&\frac{1}{Un} 
\Big[\epsilon_R+Un-\sqrt{\epsilon_R 
\left(\epsilon_R+2Un\right)}\Big], 
\nonumber\\ 
\omega_{\bf k}&=&\sqrt{\epsilon_R 
\left(\epsilon_R+2Un\right)}+i\epsilon_I. 
\end{eqnarray} 
The diagonalized Hamiltonian has the form 
\begin{eqnarray} 
{\mathcal H}_B&=&E_g 
+\sum_{{\bf k}\ne{\bf 0}}\omega_{\bf k}c_{\bf k}^Lc_{\bf k}^R, 
\nonumber\\ 
\frac{E_g}{\mathcal V}&=& 
-tn\sum_{\nu}\cosh(g_\nu a/\hbar) 
+\frac{1}{2}Un^2 
\nonumber\\ 
&&-\frac{1}{2}\frac{1}{\mathcal V}\sum_{{\bf k}\ne{\bf 0}} 
\left[\epsilon_R+Un-\sqrt{\epsilon_R 
\left(\epsilon_R+2Un\right)}\right] 
\nonumber\\ 
&&-n\frac{1}{\mathcal V} 
\sum_{{\bf k}\ne{\bf 0}}\frac{\vert V_{\bf k}\vert^2} 
{\omega_{\bf k}\omega_{-{\bf k}}}\epsilon_R. 
\label{eq:bham2} 
\end{eqnarray} 
 
From now on, we will restrict our attention to the one-dimensional case. 
For a Gaussian random potential satisfying
\begin{equation}
\overline{V_{\bf x}V_{\bf y}}=\Gamma\delta_{{\bf x},{\bf y}}
\end{equation}
and in the thermodynamic limit,
we obtain 
\begin{eqnarray}
&&\frac{\overline{{\bf J}_I}}{J_0}=\sinh(h)\Bigg\{1
+\frac{1}{n}\int_0^\pi\frac{d{\tilde k}}{\pi}
\sin^2\left({\tilde k}/2\right)
\nonumber\\
&&~~-\frac{1}{2n}\int_0^\pi\frac{d{\tilde k}}{\pi}
\frac{\left[Un+2t\cosh(h)\sin^2({\tilde k}/2)\right]\sin({\tilde k}/2)}
{\sqrt{t\cosh(h)\left[Un+t\cosh(h)\sin^2({\tilde k}/2)\right]}}
\nonumber\\
&&~~-\Gamma\int_0^\pi\frac{d{\tilde k}}{\pi}
\frac{\cosh(2h)+\cos({\tilde k})+2}{\left[
2Un\cosh(h)+t\cosh(2h)-t\cos({\tilde k})\right]^2}\Bigg\}
\nonumber\\
\end{eqnarray}
and
\begin{eqnarray} 
&&\overline{n_s}=n\cosh(h)-n_{n1}-n_{n2}, 
\nonumber\\ 
&&n_{n1}=
\int_0^\pi\frac{d{\tilde k}}{\pi} 
\Bigg\{ 
-\sin^2({\tilde k}/2)\cosh(h) 
\nonumber\\ 
&&~~+\frac{1}{2}\frac{\left[Un\cosh(h)+2t\cosh(2h)\sin^2({\tilde k}/2) 
\right]\sin({\tilde k}/2)}{\left[t 
\cosh(h)\right]^{1/2}\left[Un+t\cosh(h)\sin^2({\tilde k}/2) 
\right]^{1/2}} 
\nonumber\\ 
&&~~-\frac{1}{4}\frac{t\sinh^2(h) 
\left[Un+2t\cosh(h)\sin^2({\tilde k}/2) 
\right]^2\sin({\tilde k}/2)}{\left[t 
\cosh(h)\right]^{3/2}\left[Un+t\cosh(h)\sin^2({\tilde k}/2) 
\right]^{3/2}}\Bigg\}, 
\nonumber\\ 
&&\frac{n_{n2}}{n}=\Gamma 
\int_0^\pi\frac{d{\tilde k}}{\pi} 
\Bigg\{ 
\frac{\cosh(h) 
\left[3\cosh(2h)+\cos({\tilde k})\right]} 
{\left[2Un\cosh(h)+t\cosh(2h)-t\cos({\tilde k})\right]^2} 
\nonumber\\ 
&&-\frac{4\sinh^2(h) 
\left[Un+2t\cosh(h)\right]
\left[\cosh(2h)+\cos({\tilde k})+2\right]} 
{\left[2Un\cosh(h)+t\cosh(2h)-t\cos({\tilde k})\right]^3}\Bigg\},
\nonumber\\ 
\end{eqnarray} 
where $\tilde k=ka$. The quantity $n_{n1}$ is the contribution
to the disorder-averaged normal fluid density 
due to the breaking of the continuous translational 
symmetry in a lattice and $n_{n2}$ is the contribution
due to disorder.
 
After preforming the integrals, we find
\begin{eqnarray}
&&\frac{\overline{{\bf J}_I}}{J_0}=\sinh(h)\Bigg\{1
+\frac{1}{2n}
\nonumber\\
&&~~-\frac{1}{n\pi}\tan^{-1}\left[\sqrt{\frac{\cosh(h)}
{\tilde u}}\right]-\frac{1}{n\pi}\sqrt{\frac{\tilde u}{\cosh(h)}}
\nonumber\\
&&~~-\frac{\Gamma}{4t^2}\frac{2\cosh(h)+2\tilde u
+\tilde{u}~{\mathrm sech}^2(h)}
{\left[\tilde u+\cosh(h)\right]^{3/2}\left[\tilde u+\sinh(h)\tanh(h)
\right]^{3/2}}\Bigg\},
\nonumber\\
\end{eqnarray}
and
\begin{eqnarray} 
&&n_{n1}=
-\frac{1}{2}\cosh(h) 
\nonumber\\ 
&&+\frac{1}{\pi}\frac{1}{\cosh^2(h)}\Bigg\{
\sqrt{\tilde u\cosh(h)}\cosh(2h)
\nonumber\\
&&~+\left[\cosh(h)\cosh(2h)-\tilde u\sinh^2(h)\right]
\tan^{-1}\left[\sqrt{\frac{\cosh(h)}{\tilde u}}\right]\Bigg\}
\nonumber\\ 
&&-\frac{1}{\pi}\frac{\tanh^2(h)}{\tilde u+\cosh(h)}\Bigg\{
\sqrt{\tilde u\cosh(h)}\left[\frac{3}{2}\tilde u+\cosh(h)\right]
\nonumber\\
&&~+\left[\cosh^2(h)-\tilde u^2\right]
\tan^{-1}\left[\sqrt{\frac{\cosh(h)}{\tilde u}}\right]\Bigg\}, 
\nonumber\\ 
&&\frac{n_{n2}}{n}\nonumber\\
&&~=\frac{\Gamma}{t^2}\Bigg\{-\frac{1}{2}
\frac{1}{\left[\tilde u+\cosh(h)\right]^{1/2}\left[\tilde u+\sinh(h) 
\tanh(h)\right]^{1/2}}
\nonumber\\
&&~~~+\frac{1}{4}
\frac{\left[2\tilde u\cosh(h)+\cosh(2h)\right] 
\left[\tilde u\cosh(h)+2\cosh(2h)\right]} 
{\left[\tilde u+\cosh(h)\right]^{3/2} 
\left[\tilde u+\sinh(h)\tanh(h)\right]^{3/2} 
\cosh^2(h)} 
\nonumber\\  
&&~~~+\frac{1}{2}\frac{\left[\tilde u+2\cosh(h)\right] 
\left[2\tilde u\cosh(h)+\cosh(2h)\right]\tanh^2(h)} 
{\left[\tilde u+\cosh(h)\right]^{3/2}\left[\tilde u+\sinh(h) 
\tanh(h)\right]^{3/2}\cosh(h)} 
\nonumber\\
&&~~~-\frac{1}{8}\frac{\left[\tilde u+2\cosh(h)\right]^2\tanh^2(h)} 
{\left[\tilde u+\cosh(h)\right]^{5/2} 
\left[\tilde u+\sinh(h)\tanh(h)\right]^{5/2} 
\cosh^2(h)} 
\nonumber\\
&&~~~~~~~~\times
\left[8\tilde u^2\cosh^2(h)+8\tilde u\cosh(h)\cosh(2h)\right.
\nonumber\\
&&~~~~~~~~~~~~
\left.+\cosh(4h)+2\right]\Bigg\},
\end{eqnarray} 
where $\tilde u= Un/t$.
In the continuum limit where $a\rightarrow 0$, $ta^2\rightarrow
\hbar^2/m$ and $\Gamma a\rightarrow\gamma$, these expressions reduce
precisely to Eqs.~(\ref{eq:ic1}) and (\ref{eq:nfd1}).

\begin{figure} 
\protect\centerline{\epsfxsize=3.3in \epsfbox{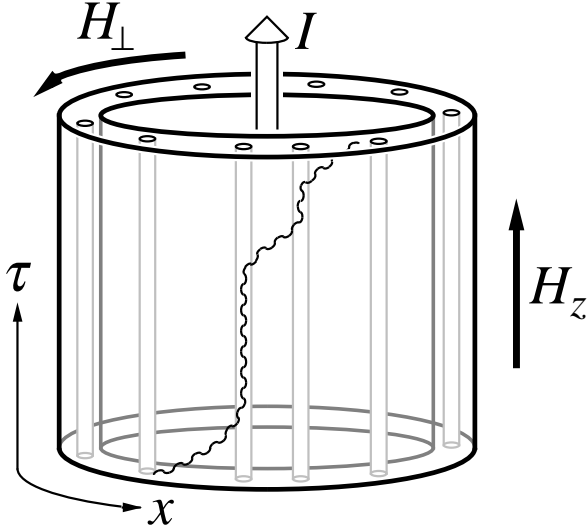}} 
\caption{One flux line (wavy curve)
induced by the field ${\bf H}_z$
and interacting with columnar defects in a cylindrical superconducting
shell. The trasnsverse component ${\bf H}_\perp$ is generated by
the current ${\bf I}$ threading the ring. In this paper, we consider
the situation where there are {\it many} interacting flux lines.}
\label{fig:pbc} 
\end{figure}

\begin{figure} 
\protect\centerline{\epsfxsize=3.3in \epsfbox{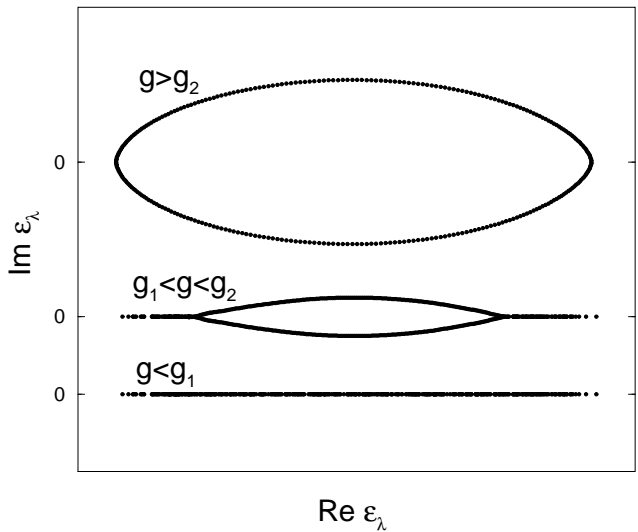}} 
\caption{Typical energy spectra of 
{\it noninteracting} bosons
in a constant imaginary vector potential $g$ and a 
random scalar potential in one dimension.}
\label{fig:hsp} 
\end{figure}

\begin{figure} 
\protect\centerline{\epsfxsize=3.3in \epsfbox{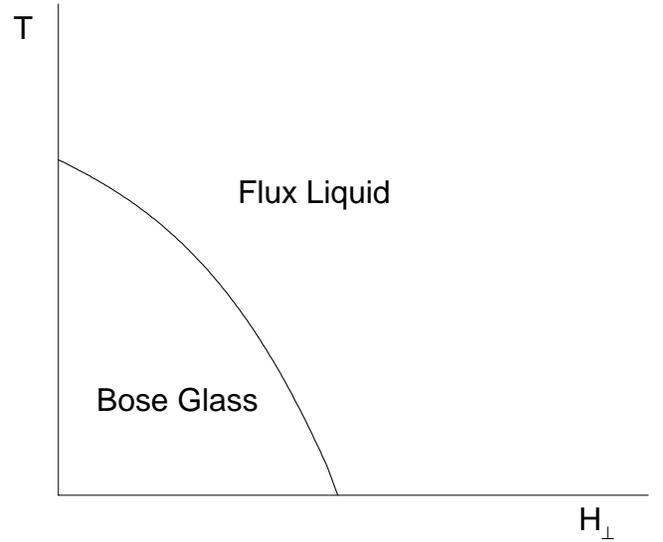}} 
\caption{{\it Schematic} phase diagram of high-temperature
superconductors with columnar defects. $H_\perp$
is the transverse component of the external magnetic field.}
\label{fig:spd} 
\end{figure}

\begin{figure}
\protect\centerline{\epsfxsize=3.3in \epsfbox{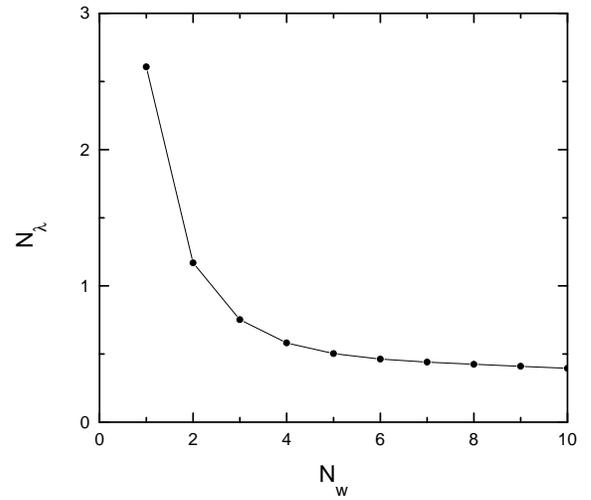}} 
\caption{Occupation number $N_\lambda=\langle b_\lambda^L
b_\lambda^R\rangle$ of the Hartree eigenstates measured in the 
Bogoliubov ground state
plotted against the winding number $N_w$ of the corresponding 
Hartree eigenfunctions. The parameters used are $Un/t=0.5$,
$\Delta/t=1$, $h=0.8$ and $L=60$.}
\label{fig:x} 
\end{figure}
  
\begin{figure} 
\protect\centerline{\epsfxsize=3.3in \epsfbox{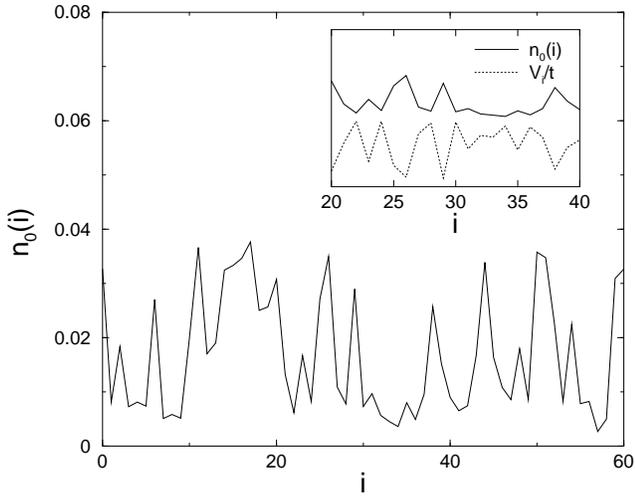}} 
\caption{Local density in the ground state $n_0(i)$
for $Un/t=0.5$, $\Delta/t=1$, $h=0.1$ and $L=60$.
Inset: Comparison of the local ground state density with 
the random potential $V_i$. $V_i$ is reduced by 30 times to make
the comparion easier. Note that maxima in $n_0(i)$ are
{\it anti}-correlated with minima in the bare potential
so that the screened potential is smoother than the bare one.}
\label{fig:cd} 
\end{figure}  
 
\begin{figure} 
\protect\centerline{\epsfxsize=3.3in \epsfbox{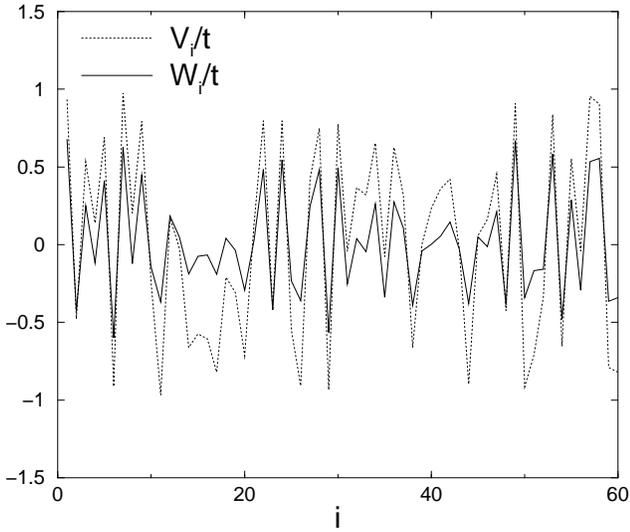}} 
\caption{Comparison of the screened random potential $W_i$ 
with the bare random potential $V_i$ for 
$Un/t=0.5$, $\Delta/t=1$, $h=0.1$ and $L=60$.}
\label{fig:srp} 
\end{figure} 

\begin{figure} 
\protect\centerline{\epsfxsize=3.3in \epsfbox{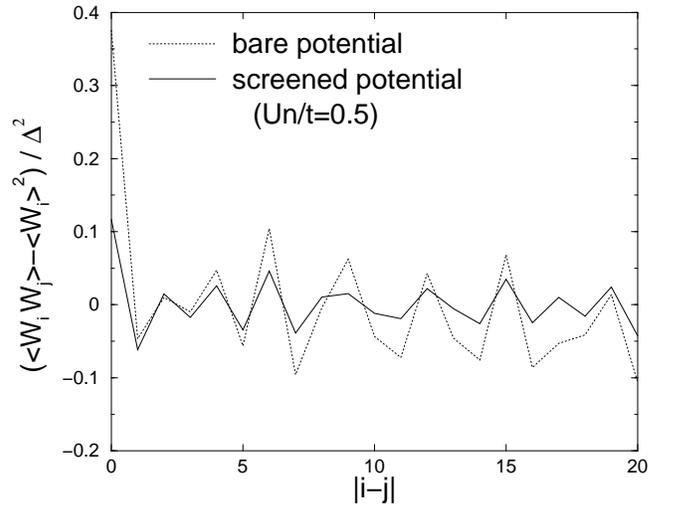}} 
\caption{Autocorrelation function of the screened 
random potential for $Un/t=0.5$, $\Delta/t=1$, $h=0.1$ and
$L=60$ compared to that of the bare random potential.}
\label{fig:cor} 
\end{figure} 
 
\begin{figure} 
\protect\centerline{\epsfxsize=3.3in \epsfbox{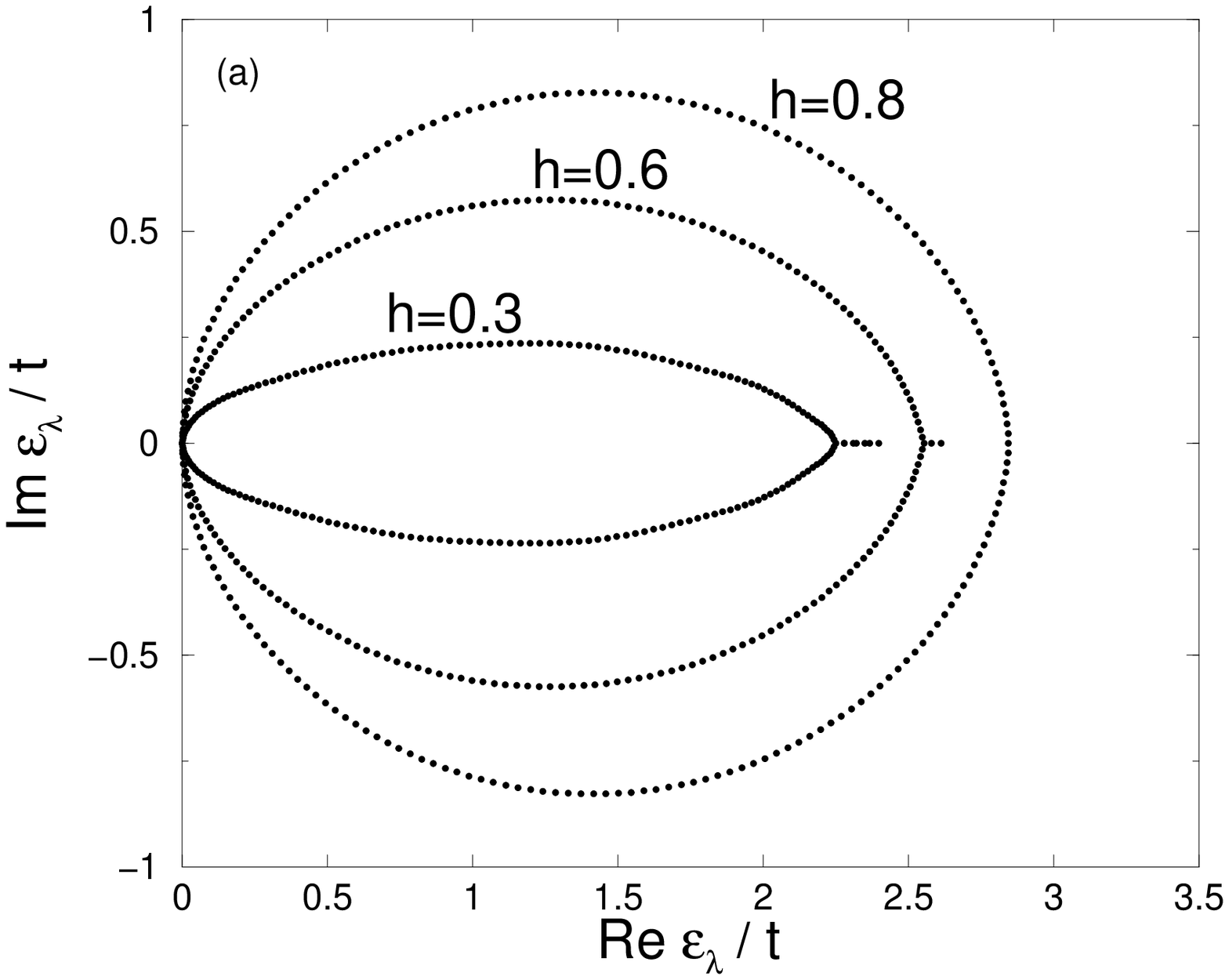}} 
\protect\centerline{\epsfxsize=3.3in \epsfbox{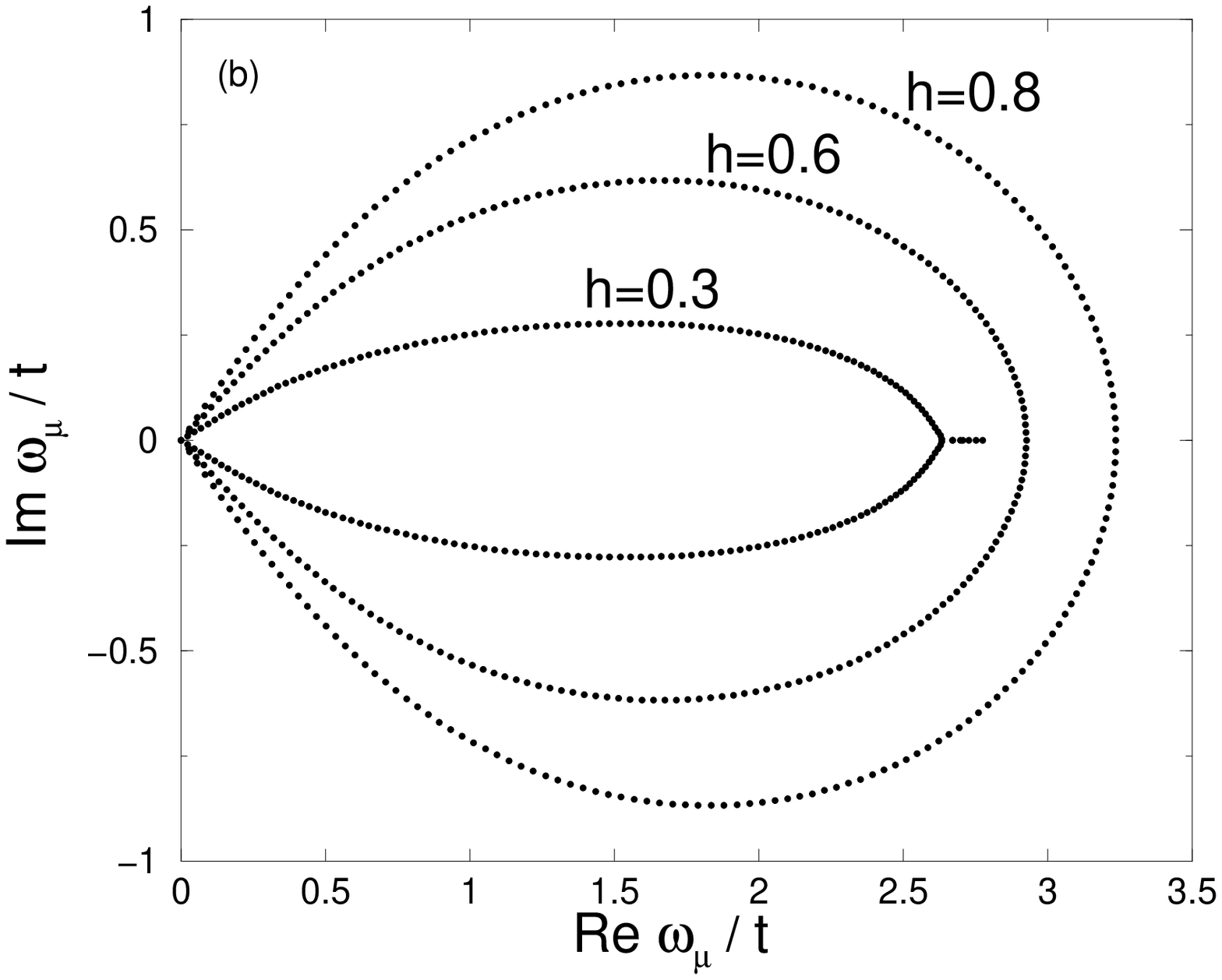}} 
\caption{(a) Hartree and (b) Bogoliubov spectra 
deep inside the superfluid
(that is, flux liquid) region for $Un/t=0.5$, $\Delta/t=1$, 
$h=0.8,0.6,0.3$ and $L=200$. The same realization of the random potential 
was used for all plots. The linear slope of the Bogoliubov
spectrum in the complex
plane near the bottom of the band reflects the non-Hermitian analogy
of the phonon part of a Bogoliubov spectrum.}
\label{fig:spec1} 
\end{figure}

\begin{figure} 
\protect\centerline{\epsfxsize=3.3in \epsfbox{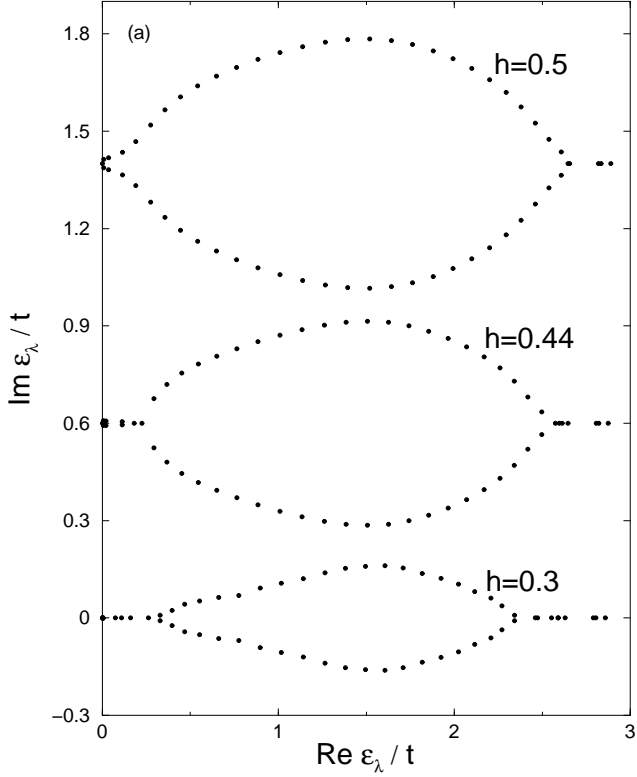}} 
\protect\centerline{\epsfxsize=3.3in \epsfbox{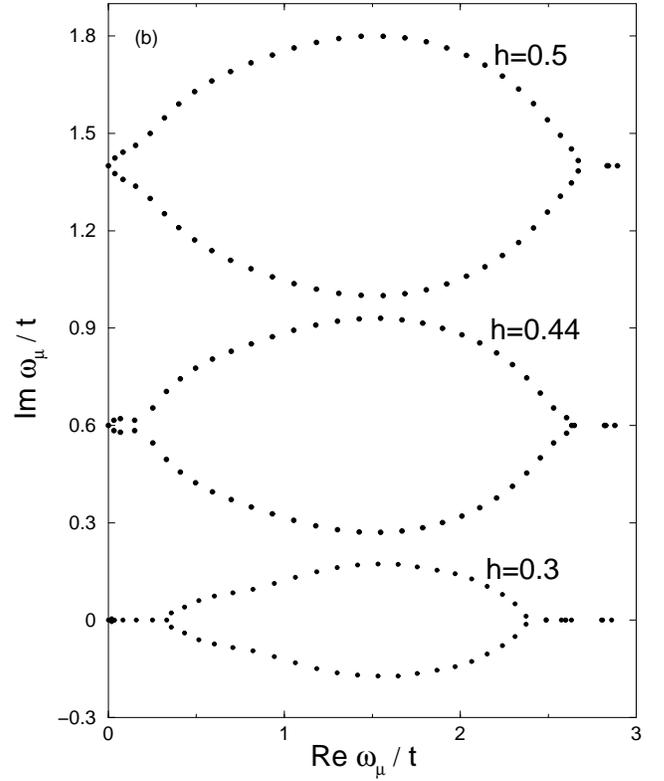}} 
\caption{(a) Hartree and (b) Bogoliubov spectra near the Bose
glass-superfluid phase boundary for $Un/t=0.05$, $\Delta/t=1$, $h=0.3,
0.44,0.5$ and $L=60$. Plots for different values of $h$ are offset for
clarity. There appears a small bubble of complex eigenvalues 
near the ground state in both spectra 
for $h=0.44$.}
\label{fig:spec2} 
\end{figure}

\begin{figure} 
\protect\centerline{\epsfxsize=3.3in \epsfbox{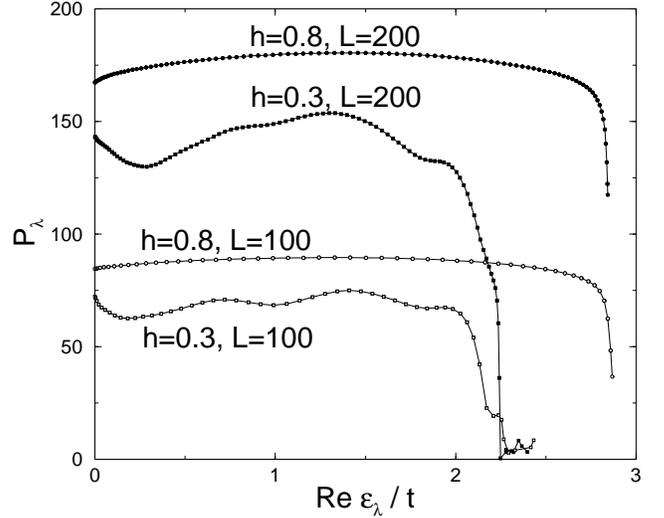}} 
\caption{Participation ratios for the Hartree eigenstates plotted 
against the real parts of their energy eigenvalues for different system
sizes. The parameters used are $Un/t=0.5$, $\Delta/t=1$, $h=0.8,0.3$
and $L=100,200$.}
\label{fig:pr} 
\end{figure}

\begin{figure} 
\protect\centerline{\epsfxsize=3.3in \epsfbox{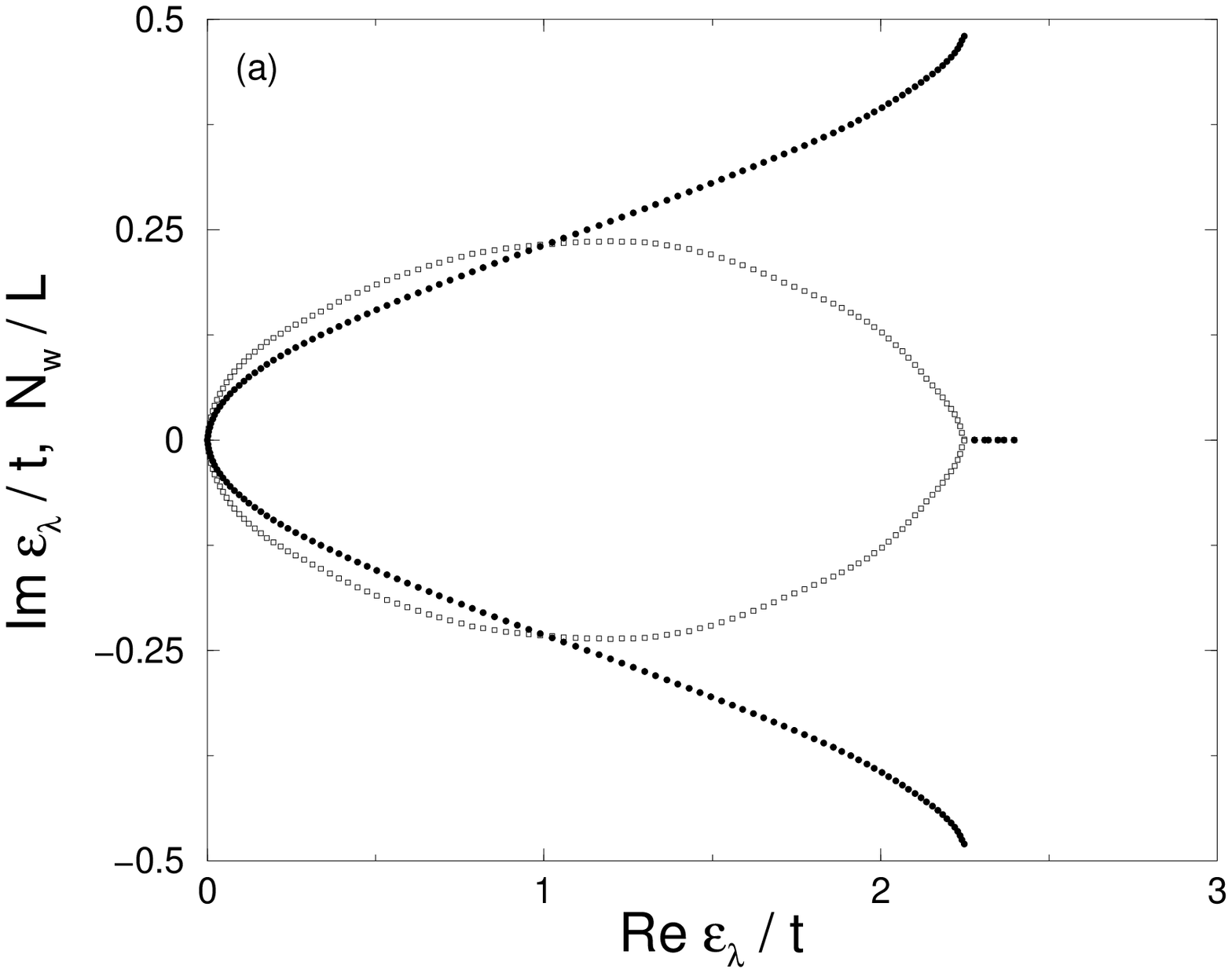}}
\protect\centerline{\epsfxsize=3.3in \epsfbox{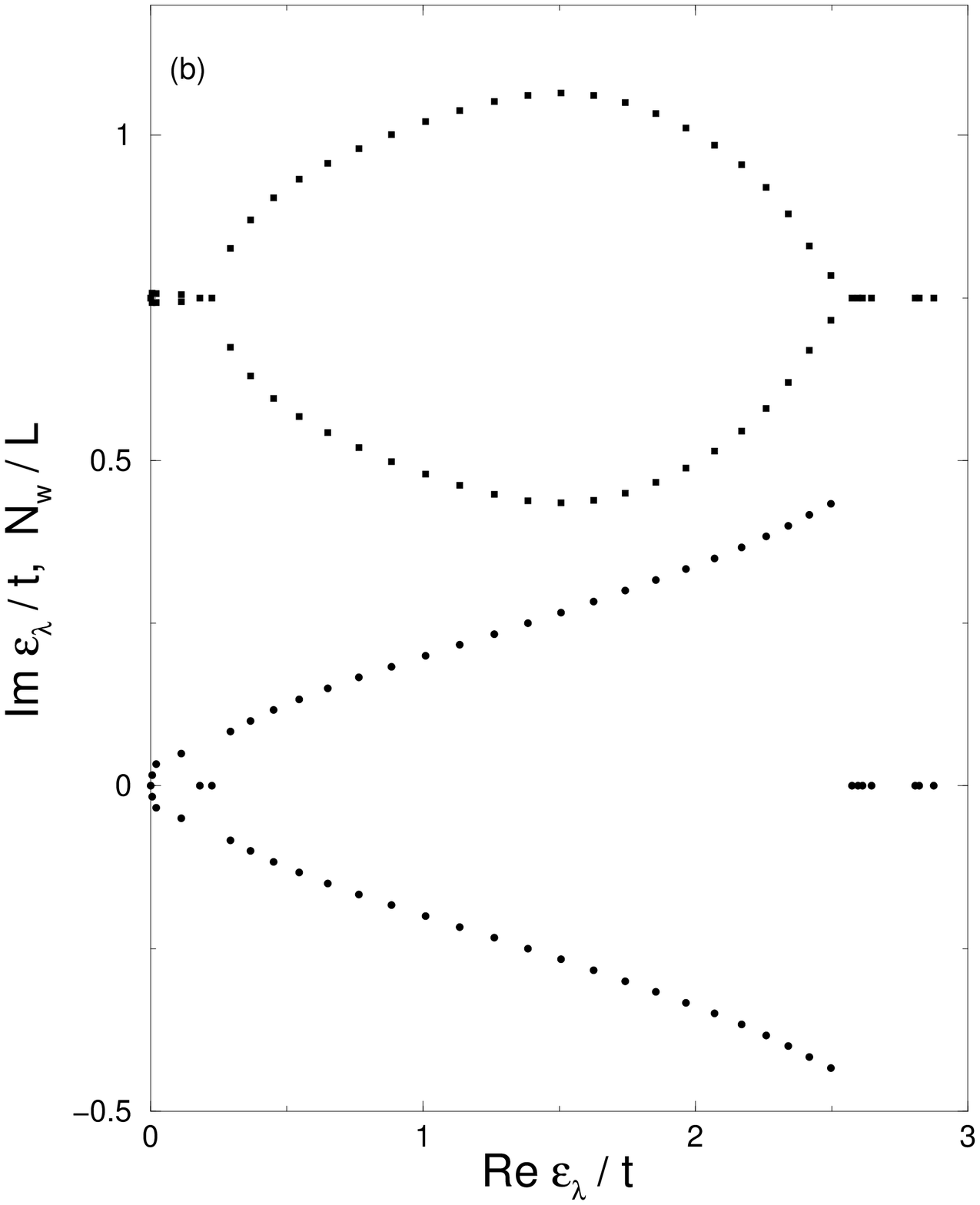}}  
\caption{Comparison of the Hartree spectrum with the winding number $N_w$ 
of the corresponding Hartree eigenfunction. (a) $Un/t=0.5$, $\Delta/t=1$, 
$h=0.3$, $L=200$. (b) $Un/t=0.05$, $\Delta/t=1$, 
$h=0.44$, $L=60$.
In (b), the plot of the spectrum is offset for
clarity. When the eigenfunction is real, we arbitrarily set $N_w=0$.}
\label{fig:wn} 
\end{figure}

\begin{figure} 
\protect\centerline{\epsfxsize=3.3in \epsfbox{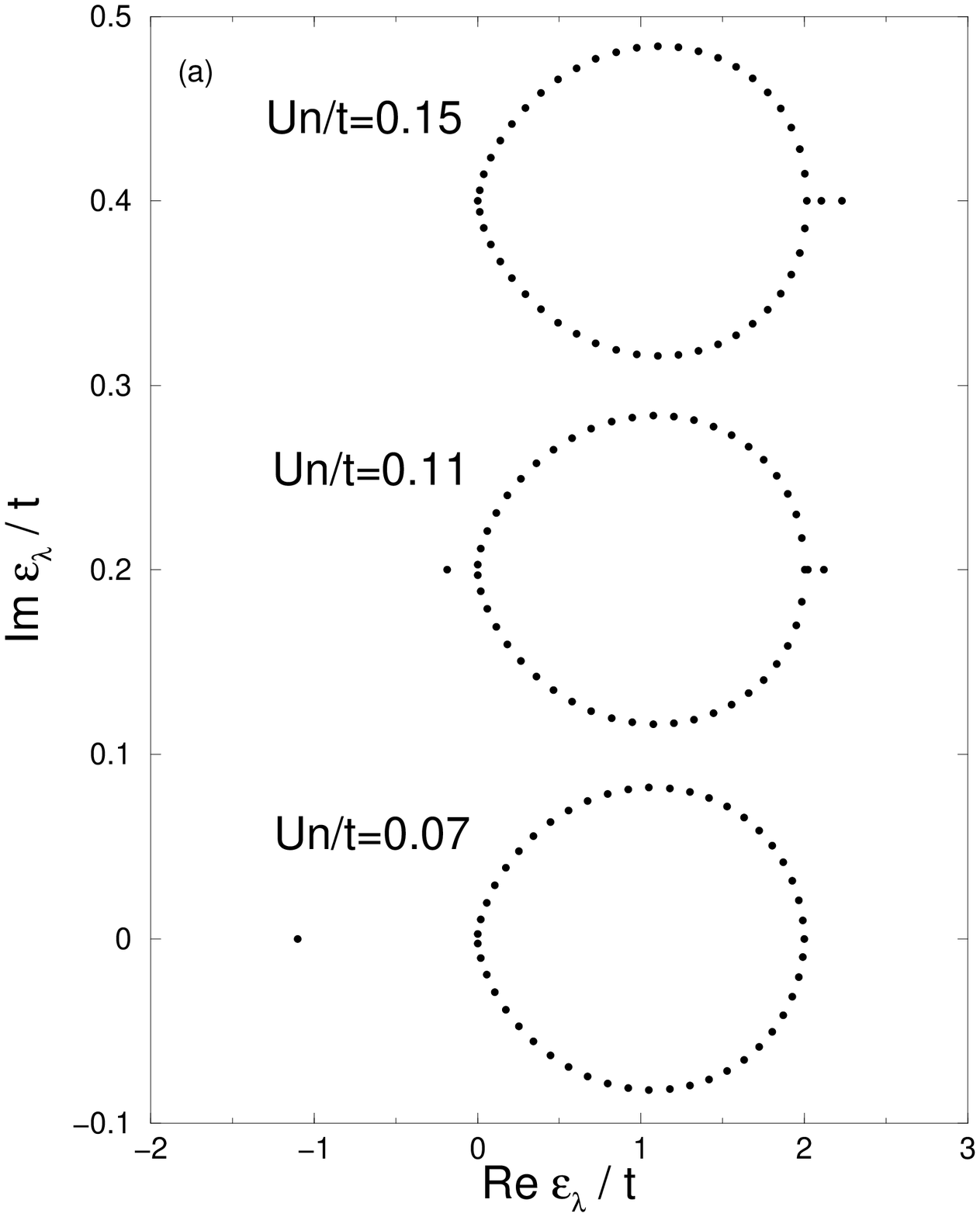}} 
\protect\centerline{\epsfxsize=3.3in \epsfbox{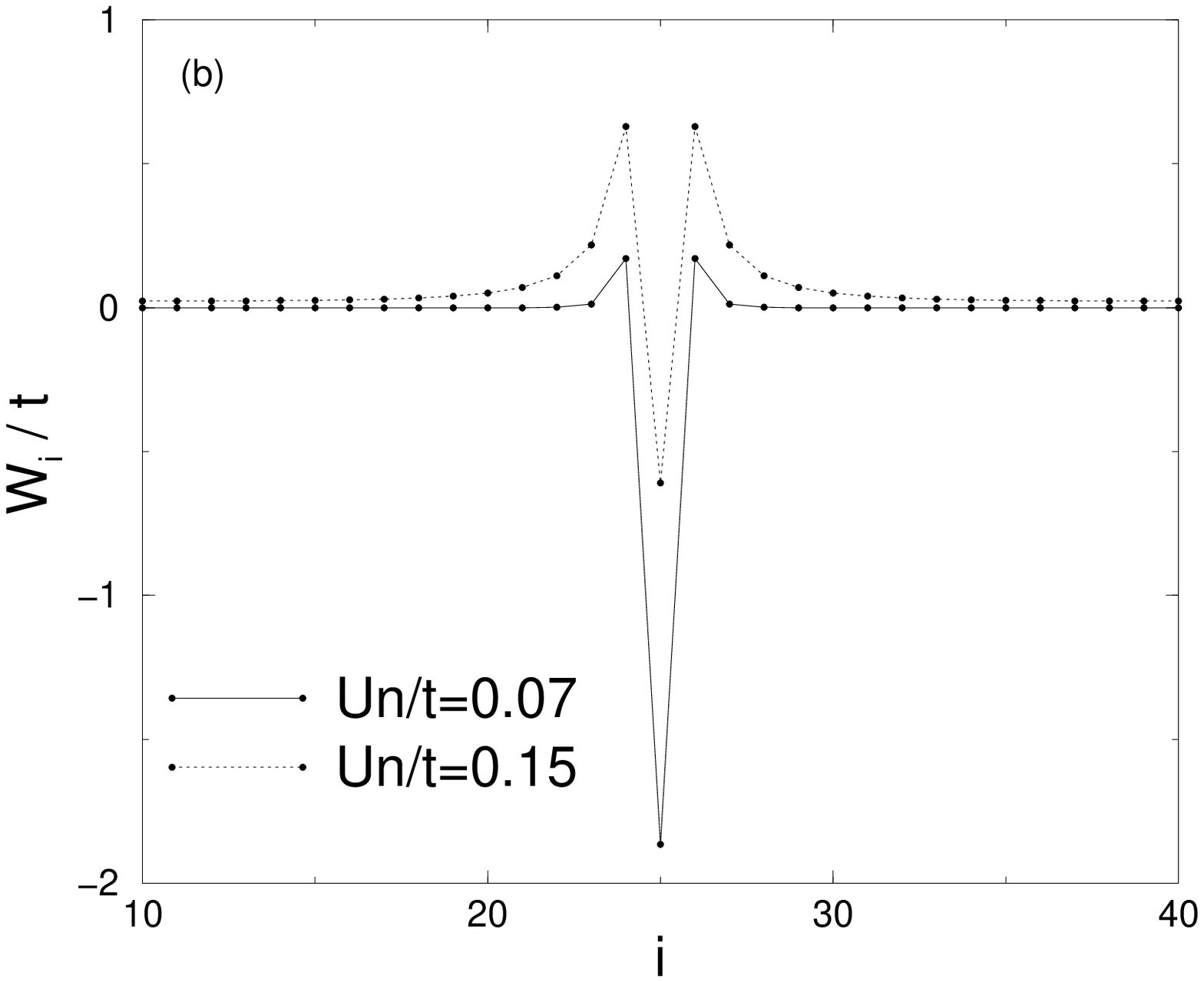}}  
\caption{(a) Hartree spectrum and (b) screened potential 
in the presence of a single attractive
impurity with $V_d/t=-5$ for $Un/t=0.07,0.11,0.15$, $h=0.1$ and $L=50$. 
The impurity is located at $i=25$.
Plots of the spectra for different values of $Un/t$ are offset for
clarity. The lowest-energy extended state is chosen to have zero energy
to align the bubbles.}
\label{fig:ospec1} 
\end{figure}

\begin{figure} 
\protect\centerline{\epsfxsize=3.3in \epsfbox{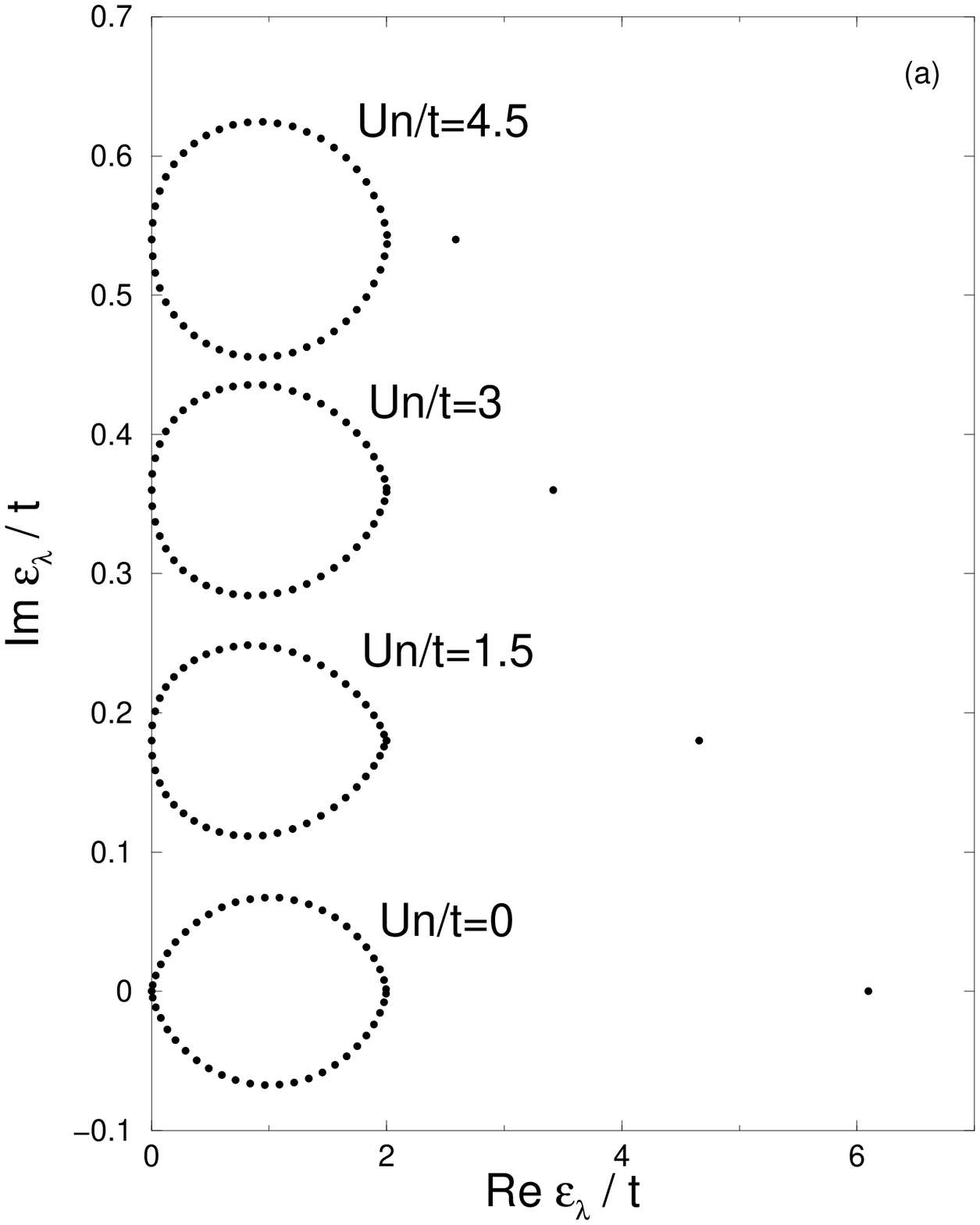}}
\protect\centerline{\epsfxsize=3.3in \epsfbox{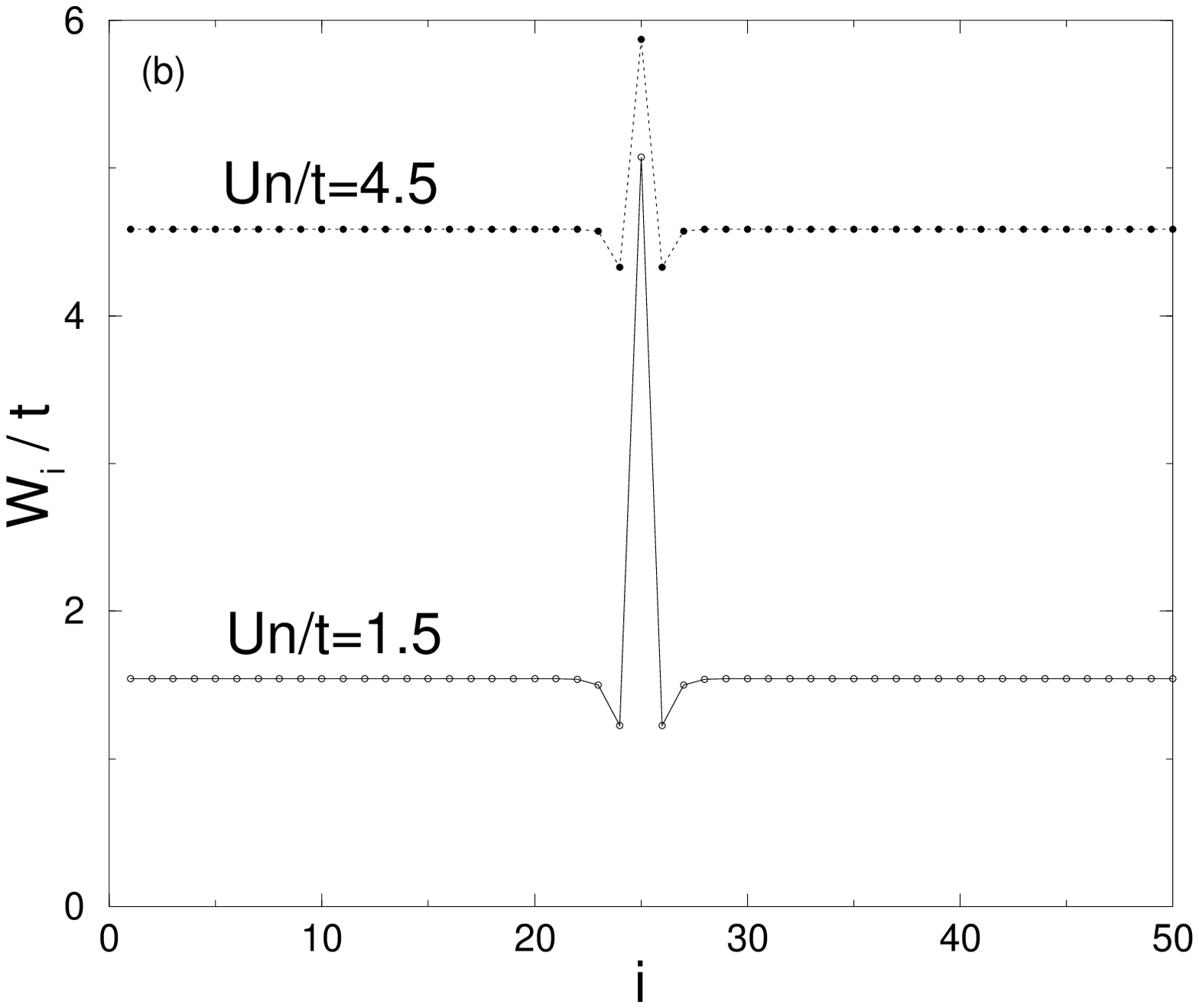}}  
\caption{(a) Hartree spectrum and (b) screened potential 
in the presence of a single repulsive
impurity with $V_d/t=5$ for $Un/t=0,1.5,3,4.5$, $h=0.1$, and $L=50$. 
The impurity is located at $i=25$.
Plots of the spectra for different values of $Un/t$ are offset for
clarity.}
\label{fig:ospec2} 
\end{figure}

\begin{figure} 
\protect\centerline{\epsfxsize=3.3in \epsfbox{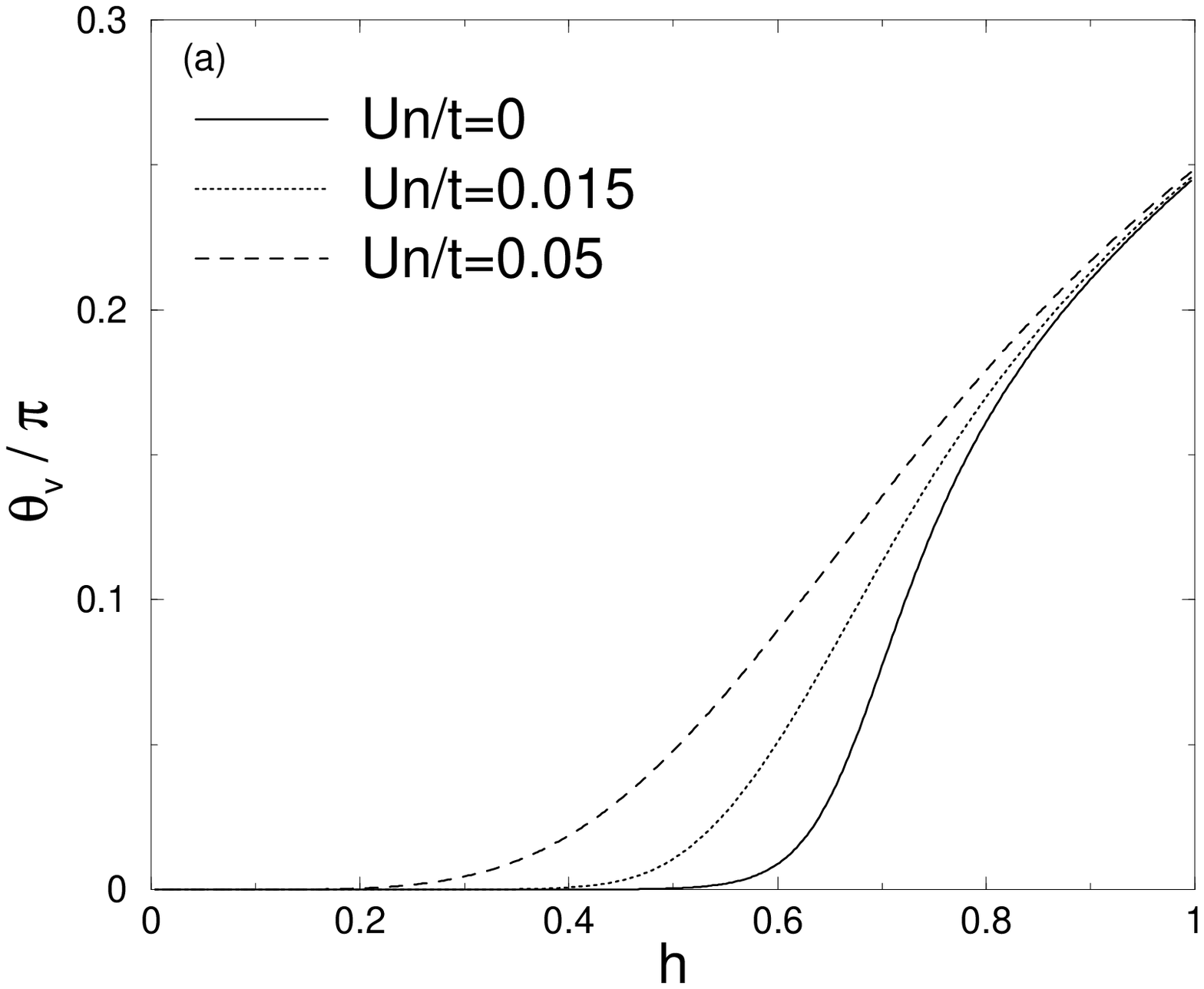}}
\protect\centerline{\epsfxsize=3.3in \epsfbox{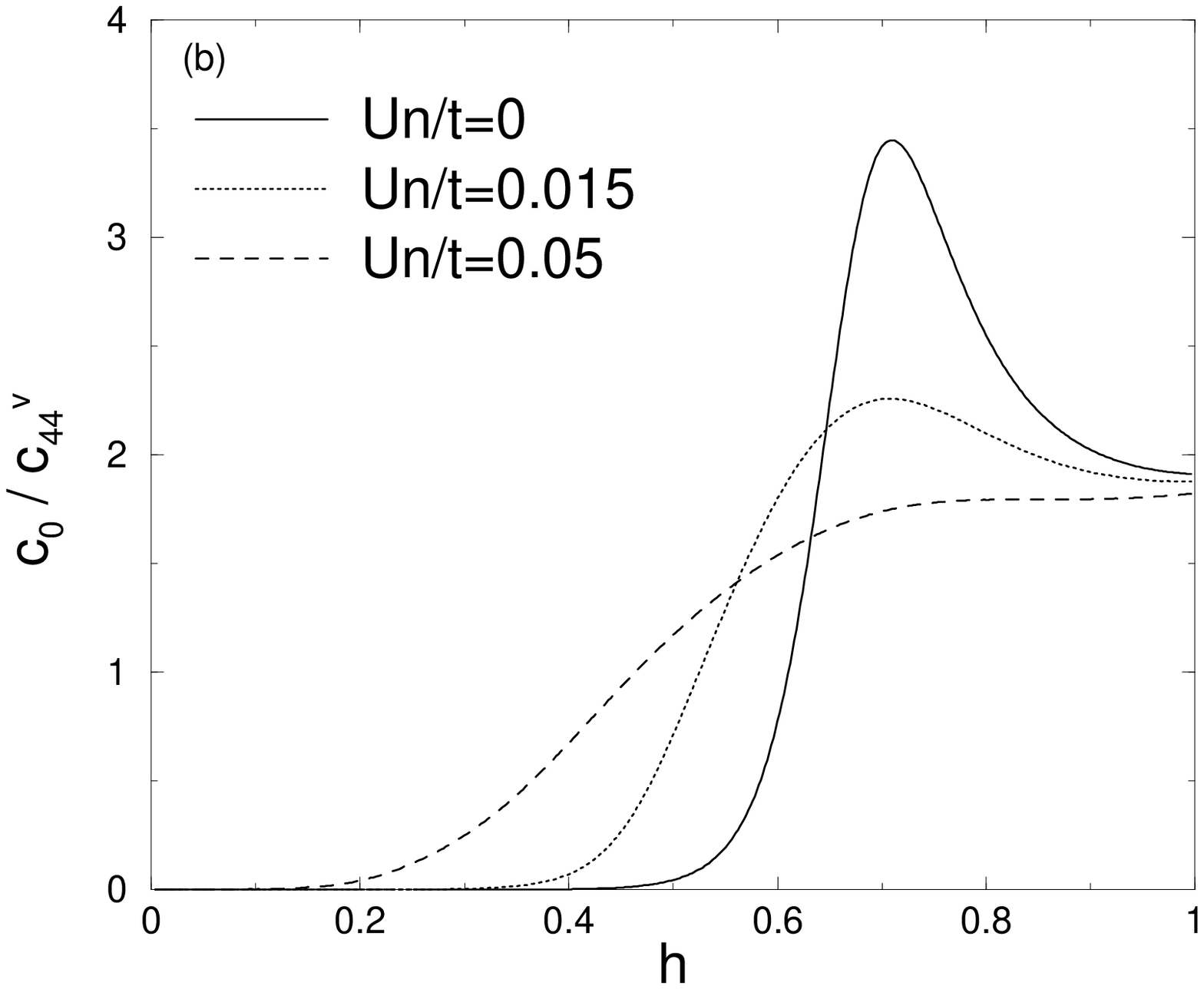}}  
\caption{(a) Vortex tilt angle $\theta_v$ and (b) normalized 
inverse tilt modulus
$c_0/c_{44}^v$ ($c_0=n\tilde\epsilon_1$) 
for $Un/t=0,0.015,0.05$, $\Delta/t=1$, $n=0.6$ and
$L=30$ as functions of $h$. }
\label{fig:tranh} 
\end{figure}

\begin{figure} 
\protect\centerline{\epsfxsize=3.3in \epsfbox{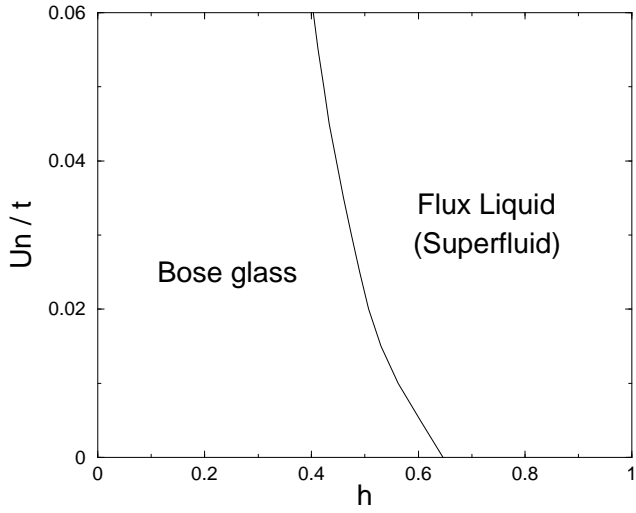}} 
\caption{Calculated phase diagram in the $U$-$h$ plane 
for a single realization of 
the random potential with $\Delta/t=1$ on a 30-site lattice.
$n$ ($=0.6$) and $t$ are fixed to constant values.
Note that interactions shrink the region of Bose glass
relative to the noninteracting case $U=0$.}
\label{fig:phase} 
\end{figure} 

\begin{figure} 
\protect\centerline{\epsfxsize=3.3in \epsfbox{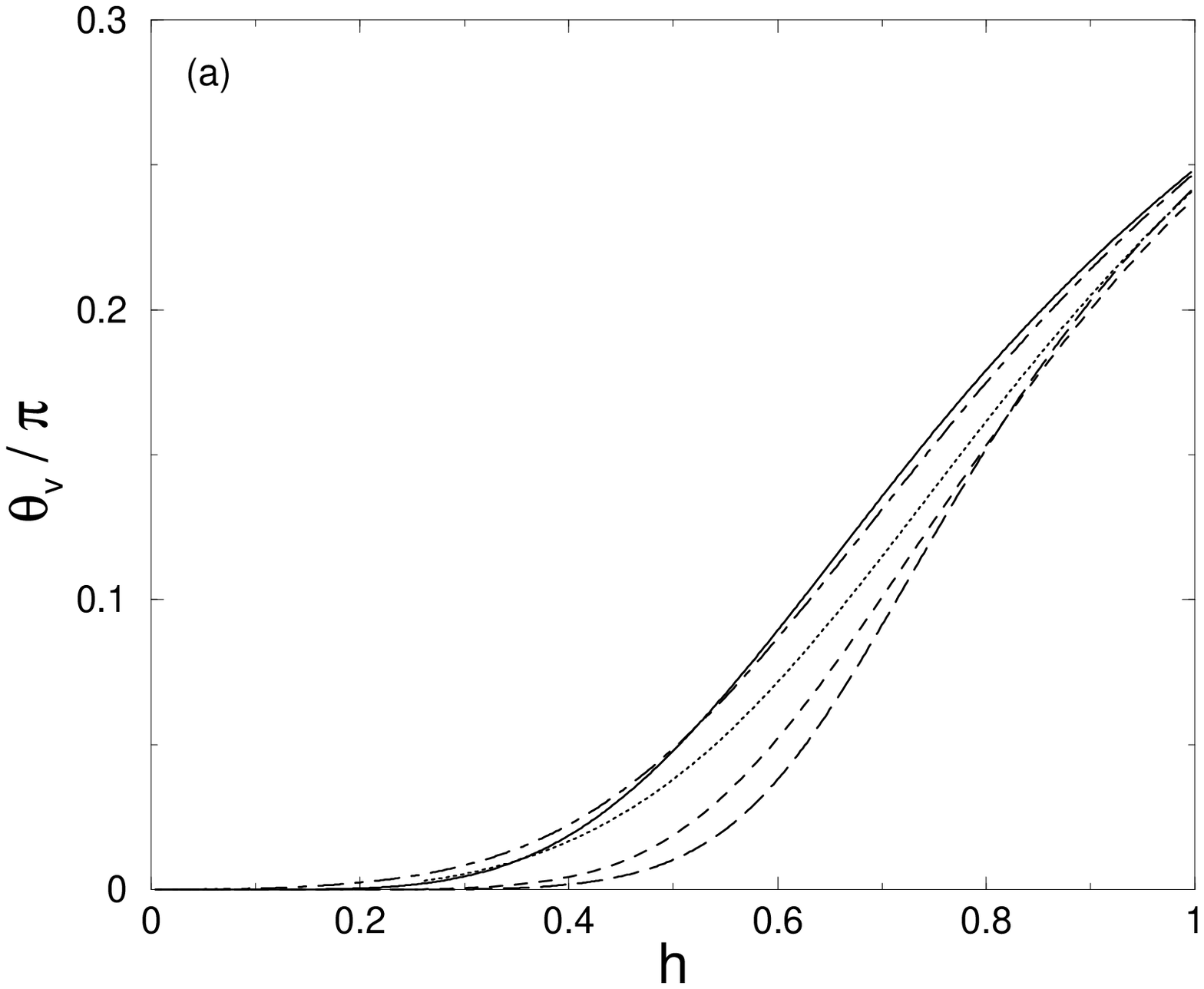}}
\protect\centerline{\epsfxsize=3.3in \epsfbox{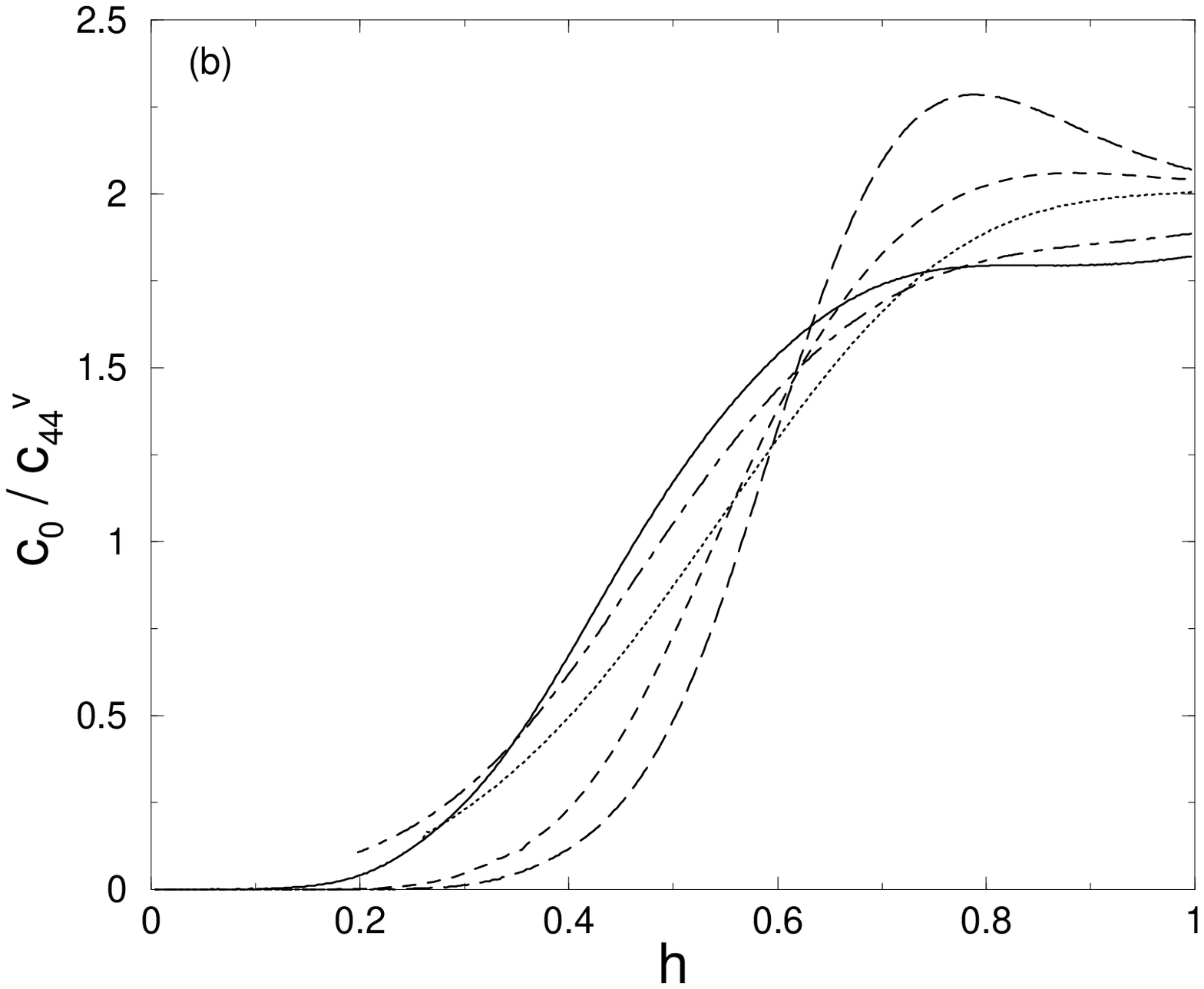}}  
\caption{Fluctuations of (a) the vortex tilt angle and (b)
the inverse tilt modulus for five different realizations 
of the random potential.
The parameters used are $Un/t=0.05$, $\Delta/t=1$, $n=0.6$ and $L=30$.}
\label{fig:fluc} 
\end{figure}

\begin{figure} 
\protect\centerline{\epsfxsize=3.3in \epsfbox{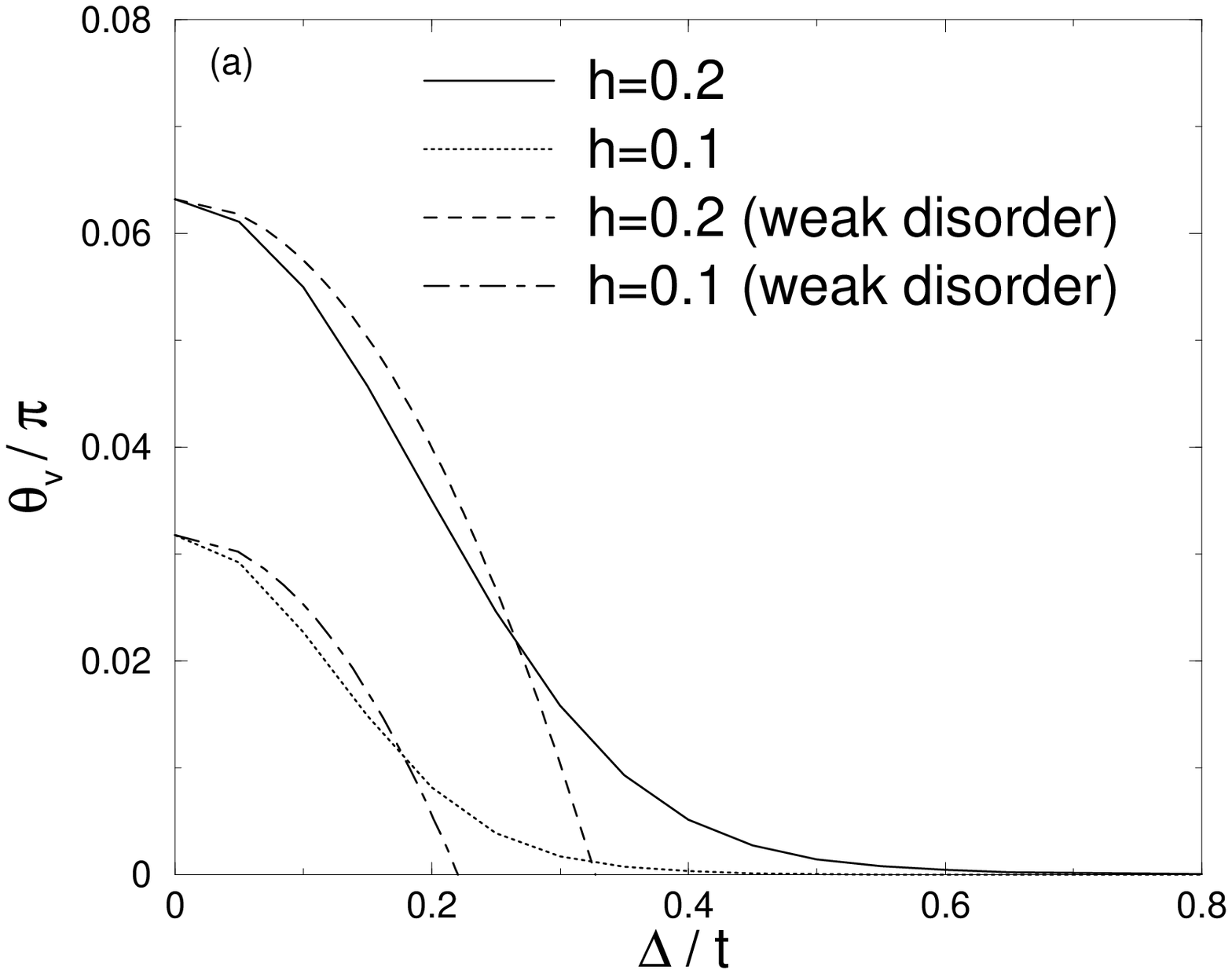}}
\protect\centerline{\epsfxsize=3.3in \epsfbox{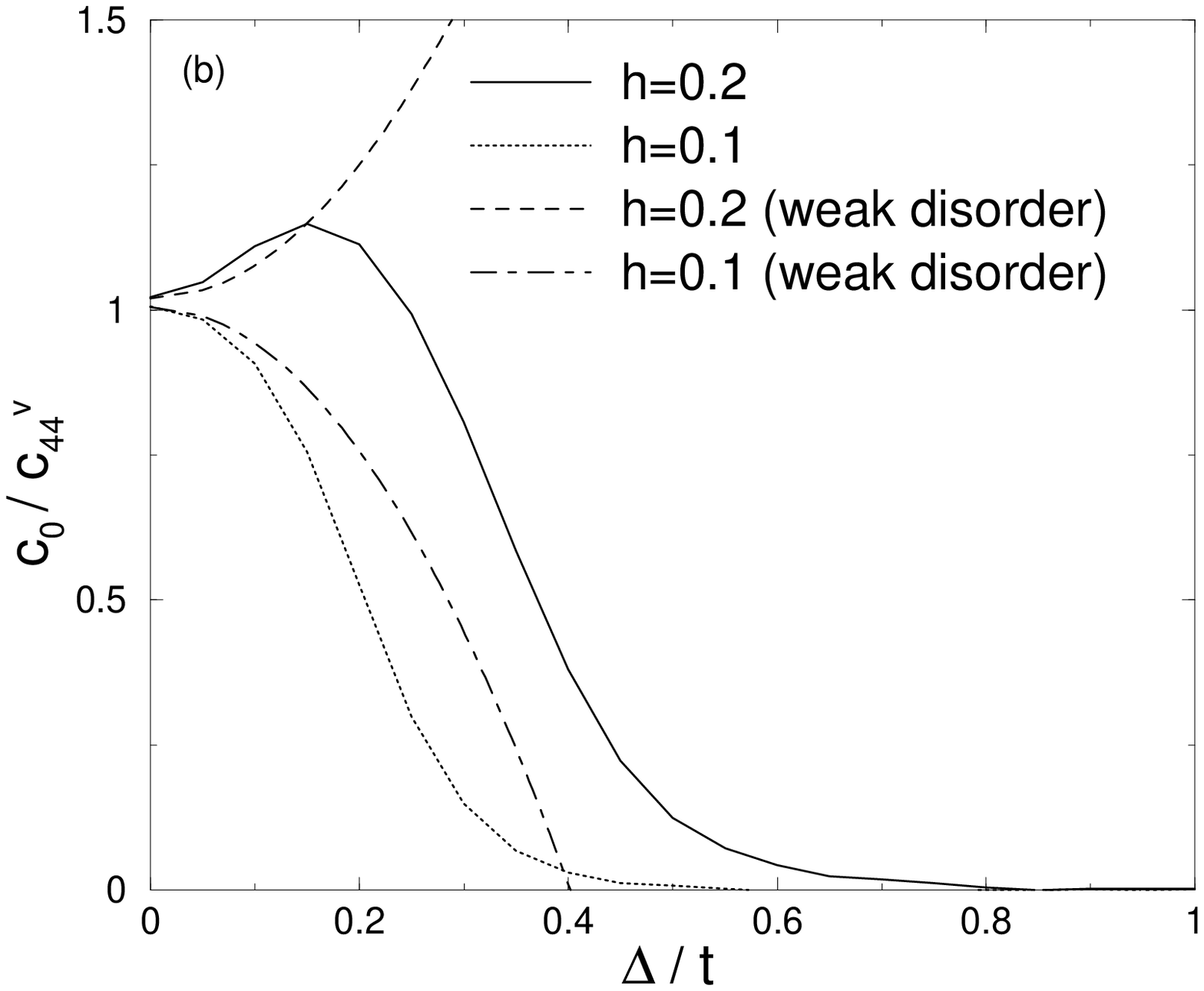}} 
\caption{(a) Vortex tilt angle and (b) 
inverse tilt modulus
for $Un/t=0.025$, $h=0.1,0.2$, $n=0.6$ and
$L=30$ as functions of $\Delta/t$. Numerical results
are compared with the analytical formulae in the weak disorder limit
derived in Appendix A. The disorder parameter in the analytical results,
$\Gamma$, is related to the parameter $\Delta$ by $\Gamma\approx
0.419\Delta^2$.}
\label{fig:trand} 
\end{figure}

\begin{figure} 
\protect\centerline{\epsfxsize=3.3in \epsfbox{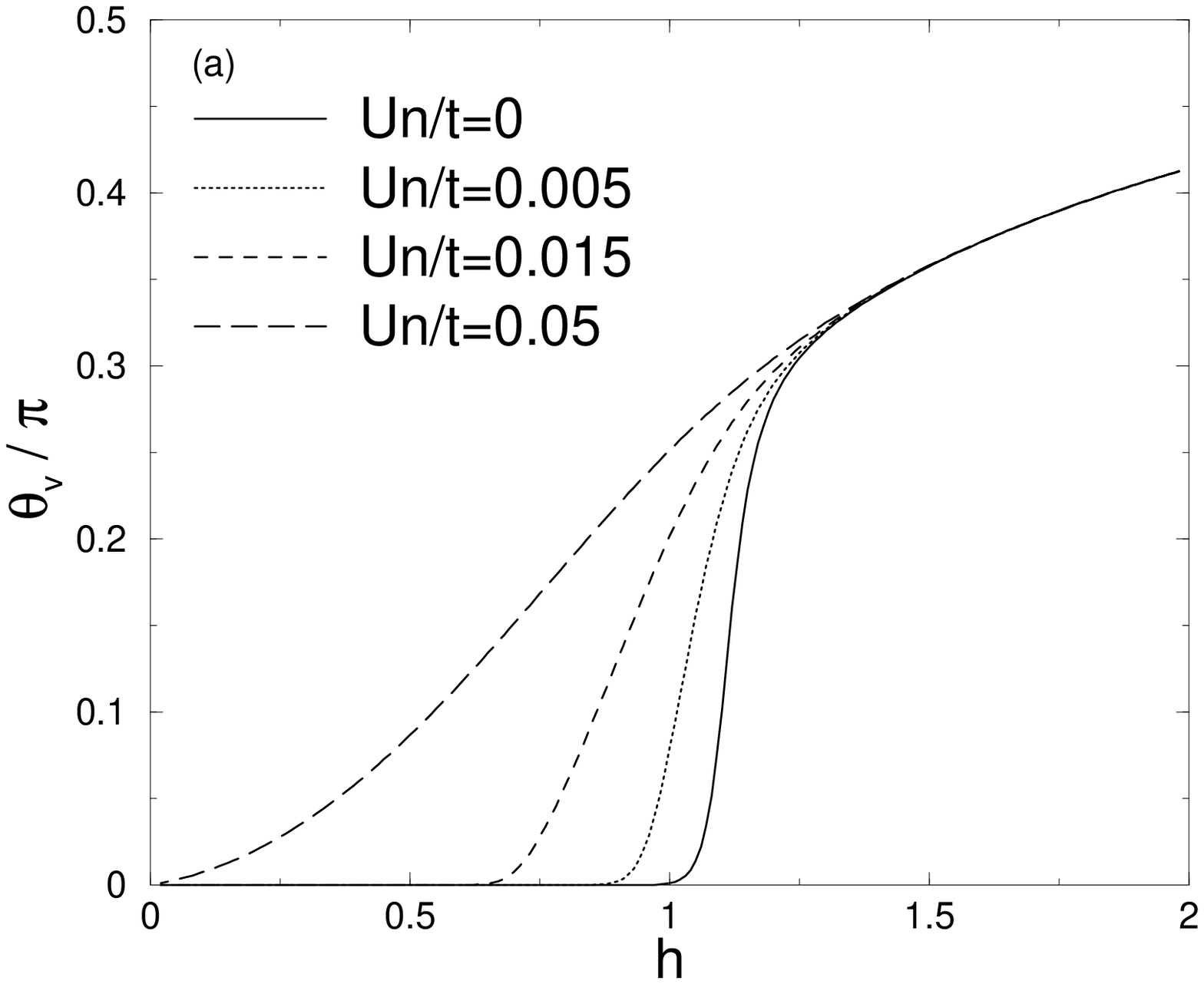}}
\protect\centerline{\epsfxsize=3.3in \epsfbox{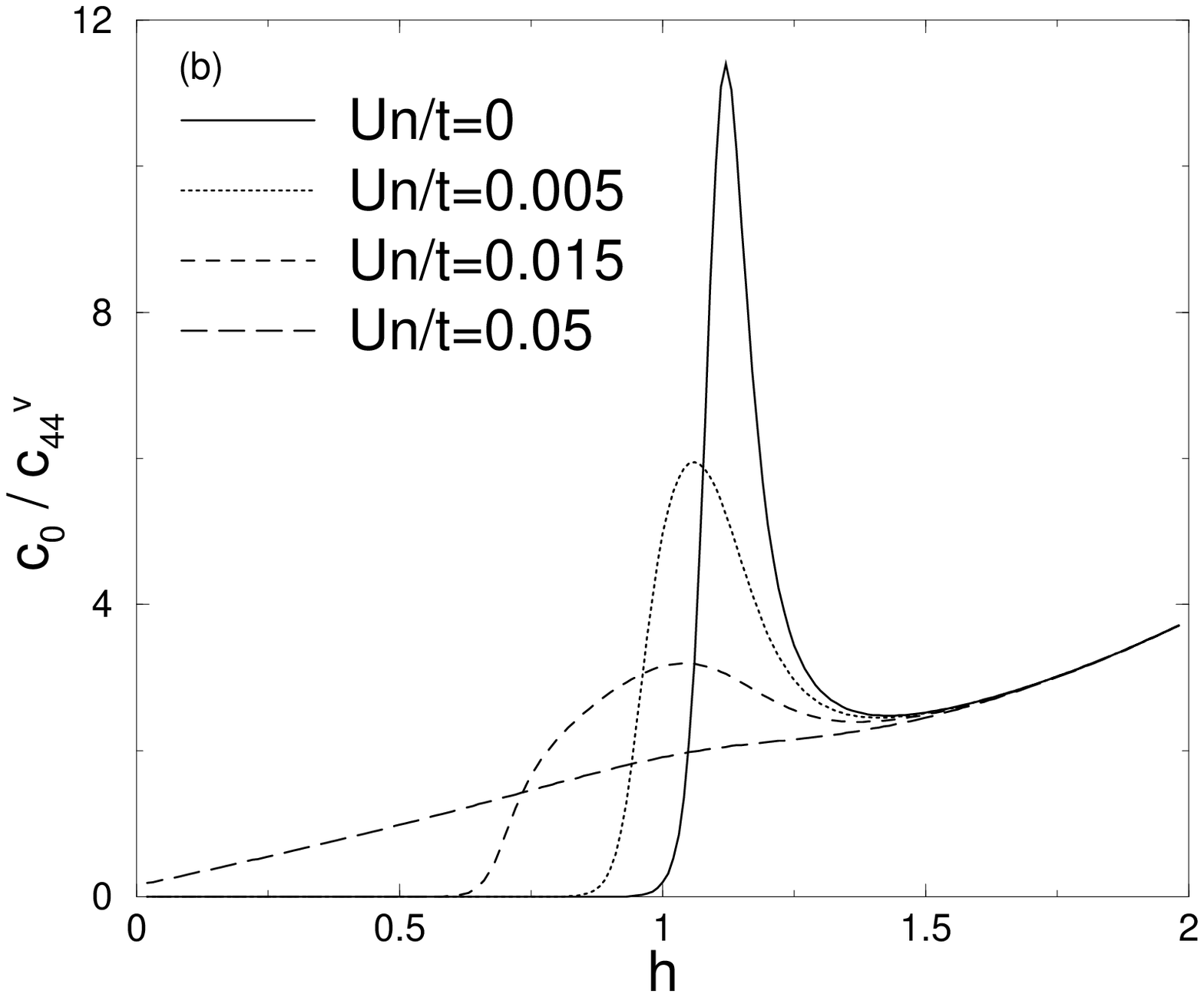}}  
\caption{(a) Vortex tilt angle and (b) 
inverse tilt modulus in the presence of a single attractive impurity
with $V_d/t=-1.5$
for $Un/t=0,0.005,0.015,0.05$, $n=0.6$ and
$L=50$ as functions of $h$.}
\label{fig:trano} 
\end{figure}

\begin{figure} 
\protect\centerline{\epsfxsize=3.3in \epsfbox{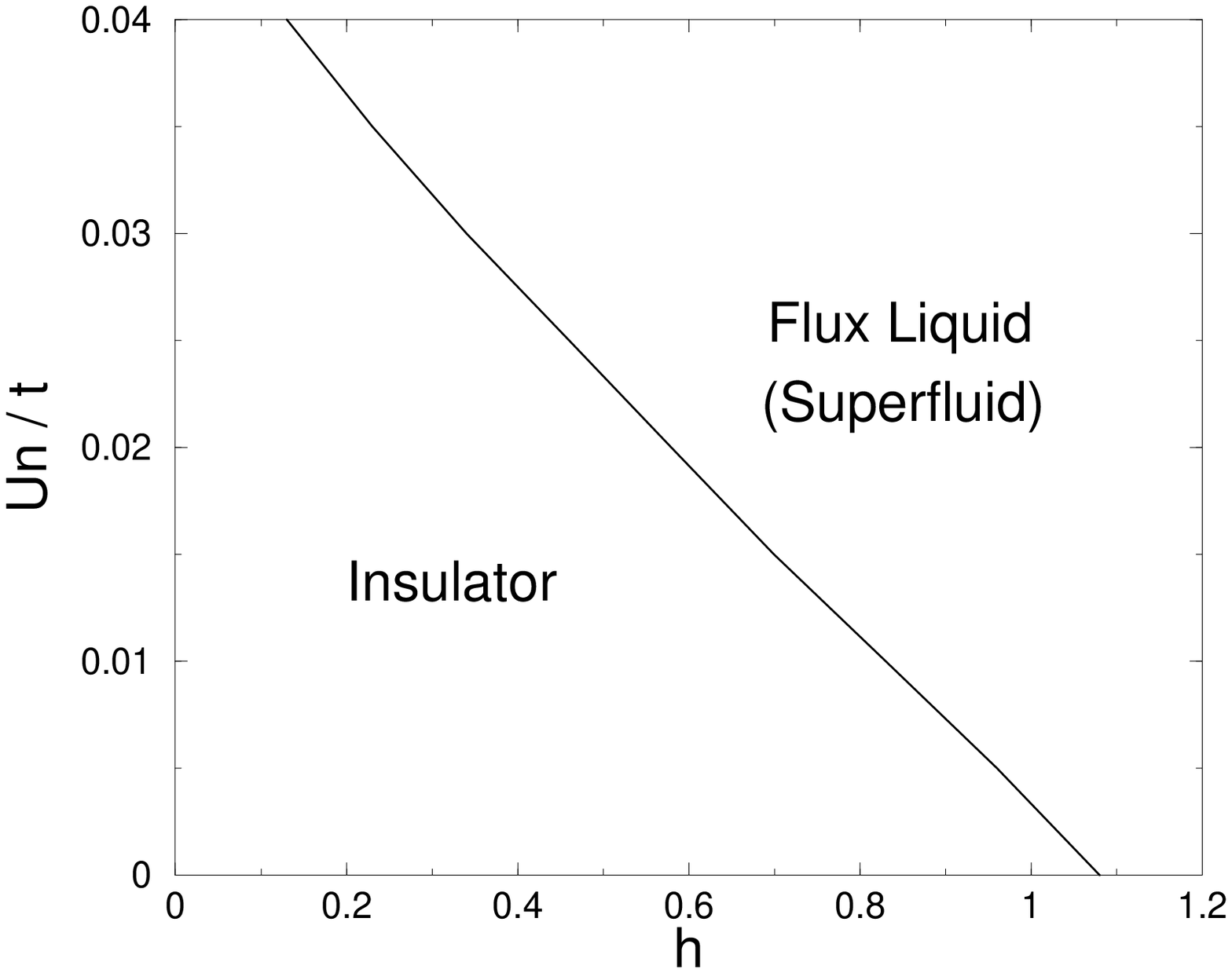}} 
\caption{Calculated phase diagram in the $U$-$h$ plane in the presence
of a single attractive impurity with $V_d/t=-1.5$ on a 50-site lattice.
$n$ ($=0.6$) and $t$ are fixed to constant values.}
\label{fig:phaseo} 
\end{figure} 
              
\begin{table} 
%\squeezetable 
\caption{Correspondence of the parameters of the vortex line system 
with those of the boson system.} 
\begin{tabular}{cc}
Vortex lines & Bosons \\ 
\tableline
$\tilde\epsilon_1$    &  $m$     \\
$k_BT$     &      $\hbar$           \\
$L_z$     &    $\beta\hbar$            \\
$\frac{B_z}{\phi_0}$   &   $n$      \\
$\frac{\phi_0H_z}{4\pi}-\tilde\epsilon_1$     &    $\mu$   \\
$\frac{\phi_0{\bf H}_\perp a}{4\pi k_BT}$   &   ${\bf h}$        \\
Flux liquid     &  Superfluid     \\
\end{tabular}
\label{table1}
\end{table}

\end{document}